\documentclass[longbibliography,twocolumn,prl,aps,superscriptaddress,showpacs,amsmath,amssymb,floatfix]{revtex4-1}
\usepackage{graphicx}
\usepackage{amssymb}
\usepackage{amsmath}
\usepackage{epsfig}
\usepackage{color}
\usepackage{tabu}
\usepackage{mathtools}
\usepackage[colorlinks,linkcolor=blue,anchorcolor=blue,citecolor=blue,urlcolor=blue]{hyperref}
\usepackage{float}

\begin{document}

\title{Many-body localization as a $d = \infty$ Anderson localization with correlated disorder}
\author{Hong-Ze Xu}
\affiliation{CAS Key Laboratory of Quantum Information, University of Science and Technology of China, Hefei, 230026, China}
\author{Shun-Yao Zhang}
\affiliation{CAS Key Laboratory of Quantum Information, University of Science and Technology of China, Hefei, 230026, China}
\author{Ze-Yu Rao}
\affiliation{CAS Key Laboratory of Quantum Information, University of Science and Technology of China, Hefei, 230026, China}
\author{Zhengwei Zhou}
\affiliation{CAS Key Laboratory of Quantum Information, University of Science and Technology of China, Hefei, 230026, China}
\affiliation{Synergetic Innovation Center of Quantum Information and Quantum Physics, University of Science and Technology of China, Hefei, Anhui 230026, China}
\affiliation{CAS Center For Excellence in Quantum Information and Quantum Physics, University of Science and Technology of China, Hefei, Anhui 230026, China}
\author{Guang-Can Guo}
\affiliation{CAS Key Laboratory of Quantum Information, University of Science and Technology of China, Hefei, 230026, China}
\affiliation{Synergetic Innovation Center of Quantum Information and Quantum Physics, University of Science and Technology of China, Hefei, Anhui 230026, China}
\affiliation{CAS Center For Excellence in Quantum Information and Quantum Physics, University of Science and Technology of China, Hefei, Anhui 230026, China}
\author{Ming Gong}
\email{gongm@ustc.edu.cn}
\affiliation{CAS Key Laboratory of Quantum Information, University of Science and Technology of China, Hefei, 230026, China}
\affiliation{Synergetic Innovation Center of Quantum Information and Quantum Physics, University of Science and Technology of China, Hefei, Anhui 230026, China}
\affiliation{CAS Center For Excellence in Quantum Information and Quantum Physics, University of Science and Technology of China, Hefei, Anhui 230026, China}
\date{\today}

\begin{abstract}
The disordered many-body systems can undergo a transition from the extended ensemble to a localized ensemble, known as many-body localization (MBL), which 
has been intensively explored in recent years. Nevertheless, the relation between Anderson 
localization (AL) and MBL is still elusive. Here we show that the MBL can be regarded as an infinite-dimensional AL with the correlated disorder in a virtual lattice.
We demonstrate this idea using the disordered XXZ model, in which the excitation of $d$ spins over the fully polarized phase can be regarded as a single-particle model in a $d$ dimensional virtual lattice. With the increasing of $d$, the system will quickly approach the MBL phase, in which the infinite-range
correlated disorder ensures the saturation of the critical disorder strength in the thermodynamic limit. From the transition from AL to MBL, the entanglement
entropy and the critical exponent from energy level statics are shown to depend weakly on the dimension, indicating that belonging to the same universal class. 
This work clarifies the fundamental concept of MBL and presents a new picture for understanding the MBL phase in terms of AL.
\end{abstract}

\maketitle

Disordered many-body localized state, known as many-body localization (MBL) \cite{basko2006metal,nandkishore2015many},
can be regarded as an elegant conceptual marriage of single-particle Anderson localization (AL) \cite{anderson1958absence,fleishman1980interactions, abrahams1979scaling} and 
many-body interaction. The origin of this MBL may be understood in terms of integrals of motion from the emergent integrability 
\cite{huse2014phenomenology}, and in some particular cases can be regarded as the fixed point of some renormalization group.  
In experiments \cite{bordia2017peridocially, schreiber2015observation,kohlert2019observation,rispoli2019quantum}, the transition between the ergodic phase and the MBL phase and their stability 
are examined on various artificially engineered platforms. The long-time slow growth of entanglement entropy (EE), an important property of the MBL phase, has 
also been observed in recent experiments by superconducting qubits \cite{xu2018emulating}, ultracold atoms \cite{lukin2019probing} and trapped ions \cite{brydges2019probing}. 

These two localizations possess several intimate similarities: both systems lack ergodicity \cite{luitz2017ergodic,deng2017many} and their energy level spacings are 
described by Wigner surmise in random matrix theory \cite{mehta2004random}. These
features make the entropy satisfy the area law, instead of volume law \cite{abanin2019many}. The MBL may be regarded as some kind of AL in the Fock space \cite{basko2006metal, bera2015many}, 
in which AL and MBL may coexist \cite{altshuler1997quasiparticle, bera2015many}. However, a large body of theoretical results are pointing to some multifaceted features 
that these two concepts are fundamentally different. For instance, MBL can be realized without 
single-particle AL \cite{lev2016many}. While any random disorder can induce AL in one dimensional model, a finite disorder strength is required for MBL \cite{luitz2015many}, which is more resemblance to three dimensional AL. It seems that the dimension is an unimportant degree of freedom in MBL since this phase can also be realized in 
quantum dot \cite{altshuler1997quasiparticle, gornyi2016many}. To date, the central concept that the relation between these two localizations is elusive 
and a  consensus definition of MBL is still lacked. 

Here we present an idea that the MBL can be regarded as an infinite-dimensional AL with the correlated disorder in a virtual lattice by the Fock states. 
We demonstrate this conclusion using the widely studied random XXZ model, in which the excitation of $d$ spins over the fully polarized state can be mapped to a $d$ dimensional tight-binding model 
with infinite-range correlated disorder. By varying $d$ from 1 to infinity, we observe a sharp transition from AL to MBL. 
The MBL is characterized by the energy level spacing ratio, with $W_c(d\rightarrow \infty) \rightarrow 3.817$ based on extrapolation
to a large $d$ limit. During the crossover from AL to MBL, the critical exponent for energy level statistics is shown to dimension independent. 
The infinite-range correlated disorder induce the saturation of $W_c$ with the increasing of dimension. The dynamics of EE is shown 
to be a smooth crossover from AL to MBL. These features indicate that the AL and MBL belong to the same universal class. Our results establish a 
clear relation between AL and MBL, which has an important application in understanding the MBL phase in various models.

\begin{figure*}
\centering
\includegraphics[clip,width=0.8\textwidth]{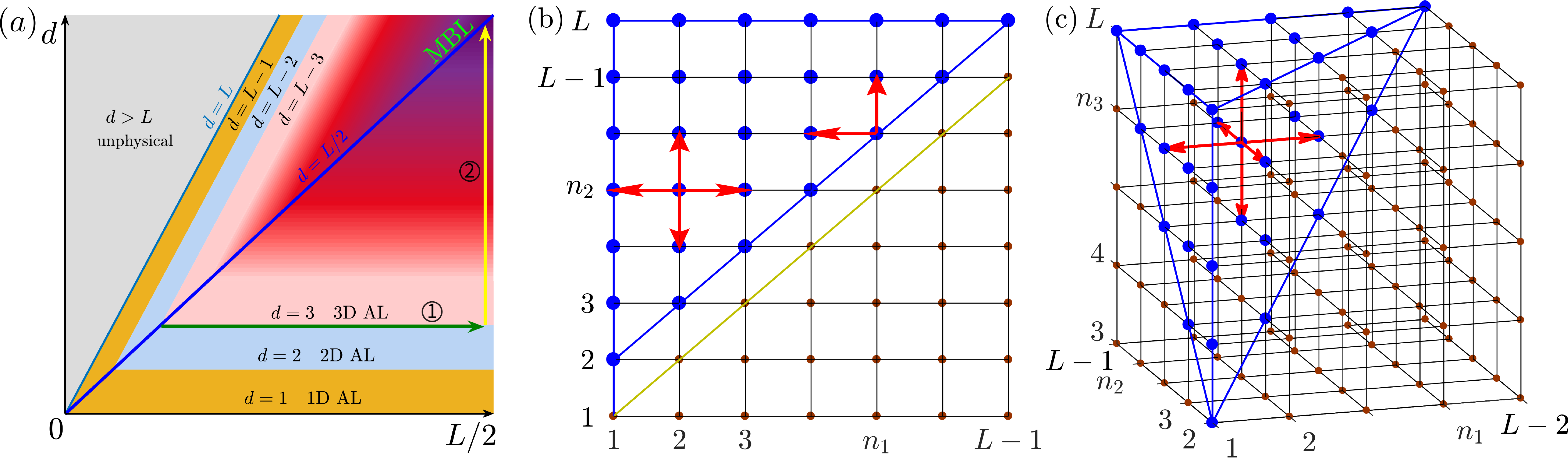}
\caption{{\bf MBL as an infinite-dimensional AL in the virtual lattice}. (a) New picture for MBL from the viewpoint of AL. The MBL phase studied in previous literature with 
$d = L/2$ spin excitation in the long-chain limit can be realized by the two steps along the horizontal and vertical arrows. (b) and (c) show the equivalent virtual lattice 
with $d = 2$ and $3$, respectively. The arrows indicate the hopping between neighboring sites, and the light points indicate the unphysical regime. This idea can be 
	straightforwardly generalized to any dimension with fermion and boson, and even models with $\mathbb{Z}_2$ symmetry \cite{supmat}.}
\label{fig-fig1}
\end{figure*}

{\it MBL and the equivalent AL in the virtual lattice}. In MBL, symmetry or conserved quantity is essential, which can decouple the Hamiltonian into different subspaces with 
MBL states to be defined therein. We consider the following widely studied disordered XXZ model \cite{agarwal2015anomalous,serbyn2014interferometric},
\begin{equation}
	H = \sum_i J(S_i^x S_{i+1}^x + S_i^y S_{i+1}^y) + J_z S_i^z S_{i+1}^z + h_i S_i^z.
	\label{eq-XXZ}
\end{equation}
We focus on $J = J_z = 1$, and $h_i \in [-W, W]$ is an uniform random potential. This model has a conserved quantity $S_\text{z} = \sum_i S_i^z = -L/2 +d$, 
where $d$ is the number of spin excitation over the fully polarized state $|\downarrow\rangle^{\otimes L}$.
It can be reduced to the interacting Fermi-Hubbard model or Bose-Hubbard model under some proper transformation, in which the $S_\text{z}$ symmetry 
is turned to the conservation of the total number of particles \cite{andraschko2014purification,luitz2017ergodic}. For this reason, our picture for MBL can also be applied to the fermion 
and boson models.  In the case of XYZ model with $\mathbb{Z}_2$ symmetry, which corresponds to the fermion and boson models with pairings, these subspaces characterized by 
$S_\text{z}$ are coupled. In this case, the subspace $S_\text{z} = 0$ (or $d = L/2$) has the largest weight. Thus even in this limit, it can still be regarded as an 
infinite-dimensional AL. These concepts are explained in details in S3 \cite{supmat}, demonstrating the generality of our picture for MBL even with other symmetries or with 
other models.

Here we will look at this phase transition in a different way (see illustration in Fig. \ref{fig-fig1} (a)). The previous literature explores the physics along
the line $d = L/2$ (thus $S_\text{z} = 0$). This limit can also be reached in a different way using two steps (see the horizontal arrow and vertical arrow). Firstly, for fixed $d$,
we can map the model to a $d$ dimensional lattice model, in which the critical disorder strength and the other quantities can be determined from the scaling of system size $L$. 
Then, we can explore these quantities by increasing the dimension $d$ to infinity. This two-step method enables us to study the crossover from AL to MBL, giving a new 
relation between them. For $d=0$, the quantum state is $|\downarrow\rangle^{\otimes L}$. For $d = 1$, the wave function can be written as
\begin{equation}
	\psi_1 =\sum_{n=1}^L c_{n} S_{n}^+|\downarrow \rangle^{\bigotimes L}\equiv \sum_{n=1}^L c_n |\phi_n\rangle,
\end{equation}
under periodic boundary condition $c_{n}=c_{n+L}$. It is important to notice that $\psi_1$ in the above equation is 
also the solution of the following tight-binding model,
\begin{equation}
	H_1 = \sum_{n=1}^L \frac{J}{2} (\hat{c}_{n}^\dagger \hat{c}_{n+1} + \text{h.c.}) +\sum_{n=1}^L (E_0 - J_z + \xi_{n} ) \hat{c}_{n}^\dagger \hat{c}_{n},
	\label{eq-H1d}
\end{equation}
where $E_0 = \frac{J_z}{4}L$, $\xi_{n} = h_{n}$ and $\hat{c}_n^\dagger$ $(\hat{c}_n)$ is the creation (annihilation) operator at site $n$. 
Noticed that a constant potential $\eta=-\sum_n^L \frac{h_n}{2}$ is discarded. The physics in this model is well known that with the random disorder the 
extended eigenstate is strictly forbidden \cite{anderson1958absence, evers2008anderson, abrahams1979scaling}, thus $W_c = 0$. The transition from localized phase to the extended 
phase can be realized in $d = 1$ only with incommensurate potential in experiments \cite{roati2008anderson,lahini2009observation}. 

\begin{figure*}
\centering
\includegraphics[width=0.8\textwidth]{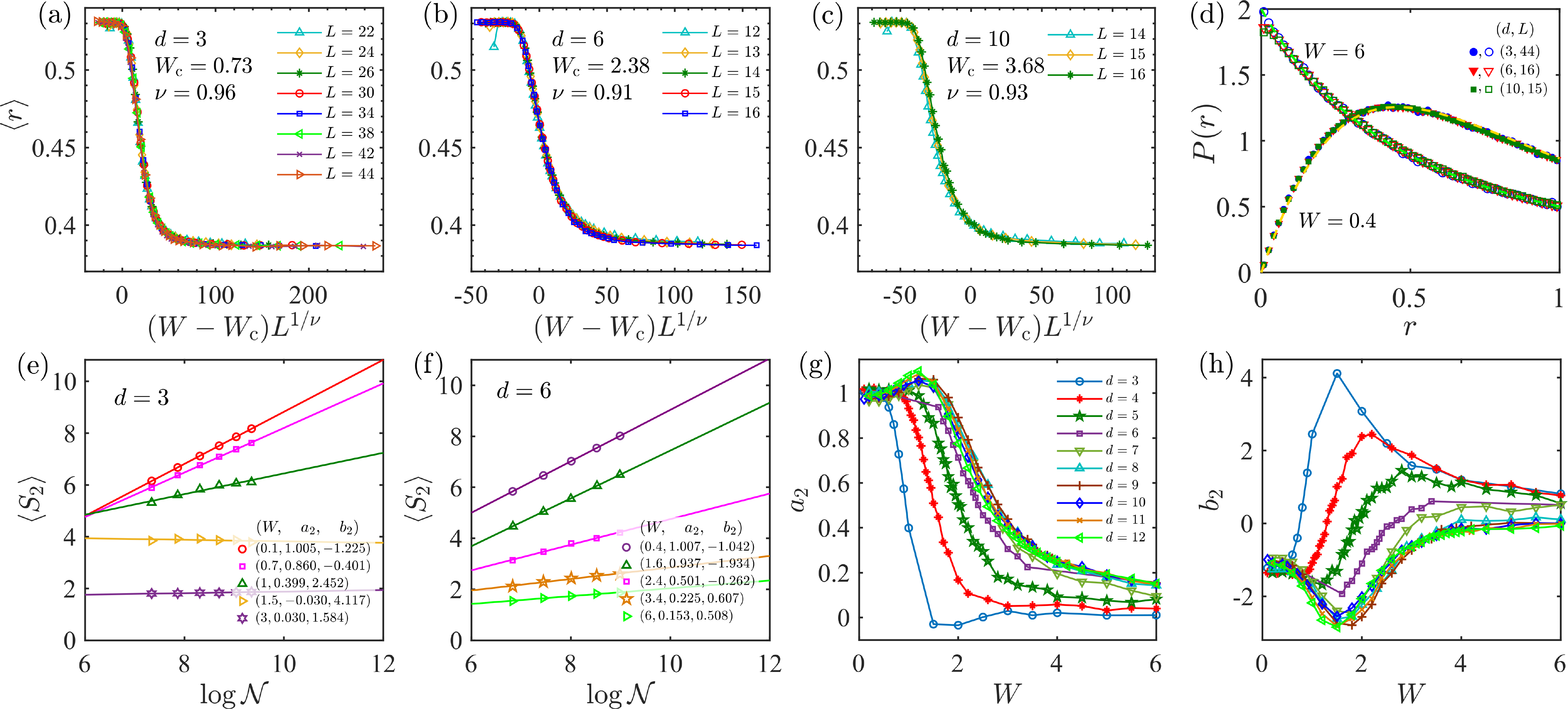}
\caption{{\bf Energy level statistics and the participation ratio of wave functions}. In the upper panel, (a) - (c) show the universal scaling of $\langle r\rangle$ using 
	Eq. \ref{eq-scalingr}, from which $W_c$ and $\nu$ are extracted. (d) show the distribution of $P_\text{GOE}$ and $P_\text{PE}$ (see Eq, \ref{eq-Pgoe}). 
	In the down panel, (e) - (f) show participated ratio $\langle S_2\rangle $ as a function of $\ln \mathcal{N}$. (g) and (h) show the fitted value of $a_2$ and 
	$b_2$ for different $d$. All data are averaged by 500 - 6000 realizations. }
\label{fig-fig2}
\end{figure*}
 
This equivalent principle is quite general and can be realized even for $d > 1$, in which the excitation of $d$ particles can be regarded as the hopping in the $d$ dimensional lattice with correlated 
disorder. To this end, let us define ${\bf n} = (n_1, n_2, \cdots, n_d)$, then the state with $d$ spin excitations can be written as
\begin{equation}
	\psi_d = \sum_{\textbf{n}}^{\mathcal{N}} c_{\textbf{n}} s_{n_1}^+ s_{n_2}^+  \cdots s_{n_d}^+ |\downarrow\rangle^{\otimes L} \equiv \sum_{\textbf{n}}^{\mathcal{N}} c_{\textbf{n}} |\phi_{\textbf{n}}\rangle.
	\label{eq-psid}
\end{equation}
with the same periodic boundary conditions $c_{n_1n_2\cdots n_d}=c_{n_2n_3\cdots n_d,n_{1+L}}$. One should be noticed that the number of lattice sites in the physics space is $\mathcal{N} = C_L^d$. 
The virtual lattice for $d = 2, 3$ are shown in Figs. \ref{fig-fig1} (b) and (c), respectively. Noticed that in the definition of this state, we have imposed
$n_i \le n_{i+1}$ for $1 \le i\le d-1$, thus the virtual lattice is only defined in a small lattice regime. This boundary maybe important for the study of the MBL phase in a finite lattice 
system, while in the large $L$ limit, its role will become negligible (S3 \cite{supmat}). The equivalent tight-binding model for $|\phi_{\bf n}\rangle$ is given by
\begin{equation}
	H_d = \sum_{\langle\textbf{n},\textbf{n}^\prime\rangle} \frac{J}{2} (\hat{c}_{\textbf{n}}^\dagger \hat{c}_{\textbf{n}^\prime} + \text{h.c.}) + [\xi_{\textbf{n}} -J_z(d-\sum_i \delta_{m_i,1}) ] \hat{c}_{\textbf{n}}^\dagger \hat{c}_{\textbf{n}},
	\label{eq-Hd}
\end{equation}
where hopping ($\langle \cdot \rangle$) is allowed between neighboring sites (see the red arrows in (b) and (c) of Fig. \ref{fig-fig1}), $m_i=n_{i+1}-n_i$, with random potential 
\begin{equation}
	\xi_{\mathbf{n}} = \sum_{i=1}^d h_{n_i},
	\label{eq-chin}
\end{equation}
and the constant potential $E_0$ and $\eta$ are discarded. We see that the many-body interaction $J_z$ and the random potential $h_i$ appear in the on-site random potential in the virtual
lattice. The correlation between the random potential is given by
\begin{equation}
	 \langle \xi_{\mathbf{n}}\xi_{\mathbf{n}^\prime}\rangle = \langle \sum_{n_i\neq n_j^\prime}^{d-k} h_{n_i} h_{n_j^\prime}\rangle + \langle \sum_{n_i = n_j^\prime}^k h_{n_i}^2\rangle  = \frac{k W^2}{3},
	 \label{eq-correlation}
\end{equation}
which indicates of infinite-range correlation with $k \le d$. We show in Ref. \cite{supmat} that the on-site random potential may also be realized by the random many-body interaction 
\cite{lev2016many, sierant2017many}, in which the single-particle AL is strictly absent. Thus our model unifies the physics of random on-site potential and random interaction.

This picture yields a new definition of MBL. It can be regarded as some kind of conceptional extension of the previous model for MBL based on the Fock state 
in the Bethe lattice (BL) \cite{de2014anderson}. In this construction, one can define $|\mathcal{G}\rangle = |N-1\rangle$ as a ground state with $N-1$ particles, then the 
$(2n-1)$-th generation of the Fock state can be defined as $ \mathcal{Y}^{i_1,\cdots,i_n}_{j_1,\cdots,j_{n-1}} 
= \hat{c}_{i_1}^\dagger \cdots \hat{c}_{i_n}^\dagger \hat{c}_{j_1} \cdots \hat{c}_{j_{n-1}}|\mathcal{G}\rangle$ 
\cite{altshuler1997quasiparticle,supmat}. These generations can form a lattice structure (see S6 \cite{supmat}), which under some approximation can be simplified
to the cycle-free BL. Recent investigation has revealed the intimate relation between AL in the BL with on-site random potential and MBL \cite{saberi2015recent}; 
nevertheless, it fails to describe the relationship between the AL and MBL. In our model, the Fock state is used, however, without the need of ground state $|\mathcal{G}\rangle$; 
and the node is arranged in a $d$ dimensional virtual lattice, thus the crossover from AL to MBL can be realized by increasing of dimension $d$.

{\it Phase transition and crossover from AL to MBL}. To characterize the transition from the ergodic phase to the ergodic breaking phase in the $d$ dimensional virtual lattice, we use the widely 
explored $r$-statistics $r =  \min (s_{i+1},s_i) / \max (s_{i+1},s_i)$, with $s_i = E_{i+1}-E_{i}$ ($\{E_i\}$ sorted in ascending order) \cite{atas2013distribution}. 
In the ergodic phase by the Gaussian orthogonal ensemble (GOE), $\langle r\rangle_{\text{GOE}}=0.5307$. In contrast, in the MBL phase with Poisson ensemble (PE), $\langle r\rangle_{\text{PE}}=0.3863$. 
Our results for $d = 3$, 6, 10 are presented in Figs. \ref{fig-fig2} (a) - (c); more details can be found in Ref. \cite{supmat}. In these figures, we have adopted the scaling 
ansatz 
\cite{luitz2015many} 
\begin{equation}
	\langle r\rangle = f[(W-W_c) L^{1/\nu} ].
	\label{eq-scalingr}
\end{equation}
With this method, we can obtain $W_c$ and its corresponding critical exponent $\nu$ for different dimensions. 
In Fig. \ref{fig-fig2} (d), these two curves collapse to the universal form in the two ensembles \cite{atas2013distribution}
\begin{equation}
	P_\text{GOE}(r) = \frac{27(r+r^2)}{4(1+r+r^2)^{5/2}}, \quad P_\text{PE}(r)=\frac{2}{(1+r)^2}.
	\label{eq-Pgoe}
\end{equation}

\begin{figure} 
\centering
\includegraphics[width=0.4\textwidth]{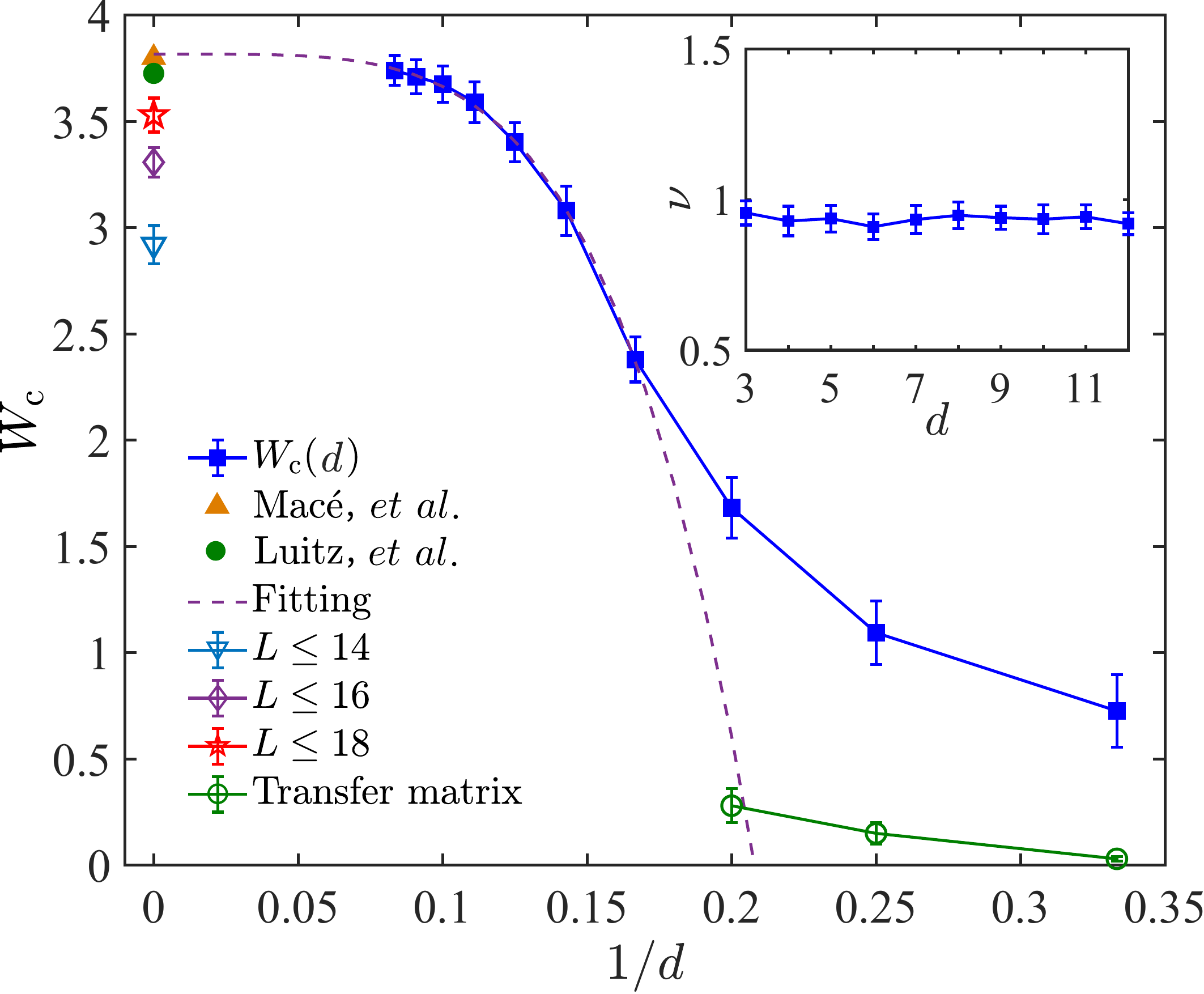}
	\caption{{\bf Smooth crossover from AL to MBL}. The relation between $1/d$ and $W_c$ is presented by blue squares. The dashed line is the best fitting using Eq. \ref{eq-crossoverWc}, yielding $W_c(\infty)=3.817$, $\alpha=-3794$ and $\beta=4.396$. Here $W_c$ is slightly greater than $W_c = 3.72$ by Luitz {\it et. al.} \cite{luitz2015many} and $W_c = 3.8$	by Mac\'e {\it et. al.} \cite{mace2018multifractal,pietracaprina2018shift}.  The green circles are results from the transfer matrix (see Ref. \cite{supmat}). The inset shows the corresponding critical exponents $\nu$, which are almost independent of dimension $d$.}
\label{fig-fig3}
\end{figure}

\begin{figure}
\centering
\includegraphics[width=0.36\textwidth]{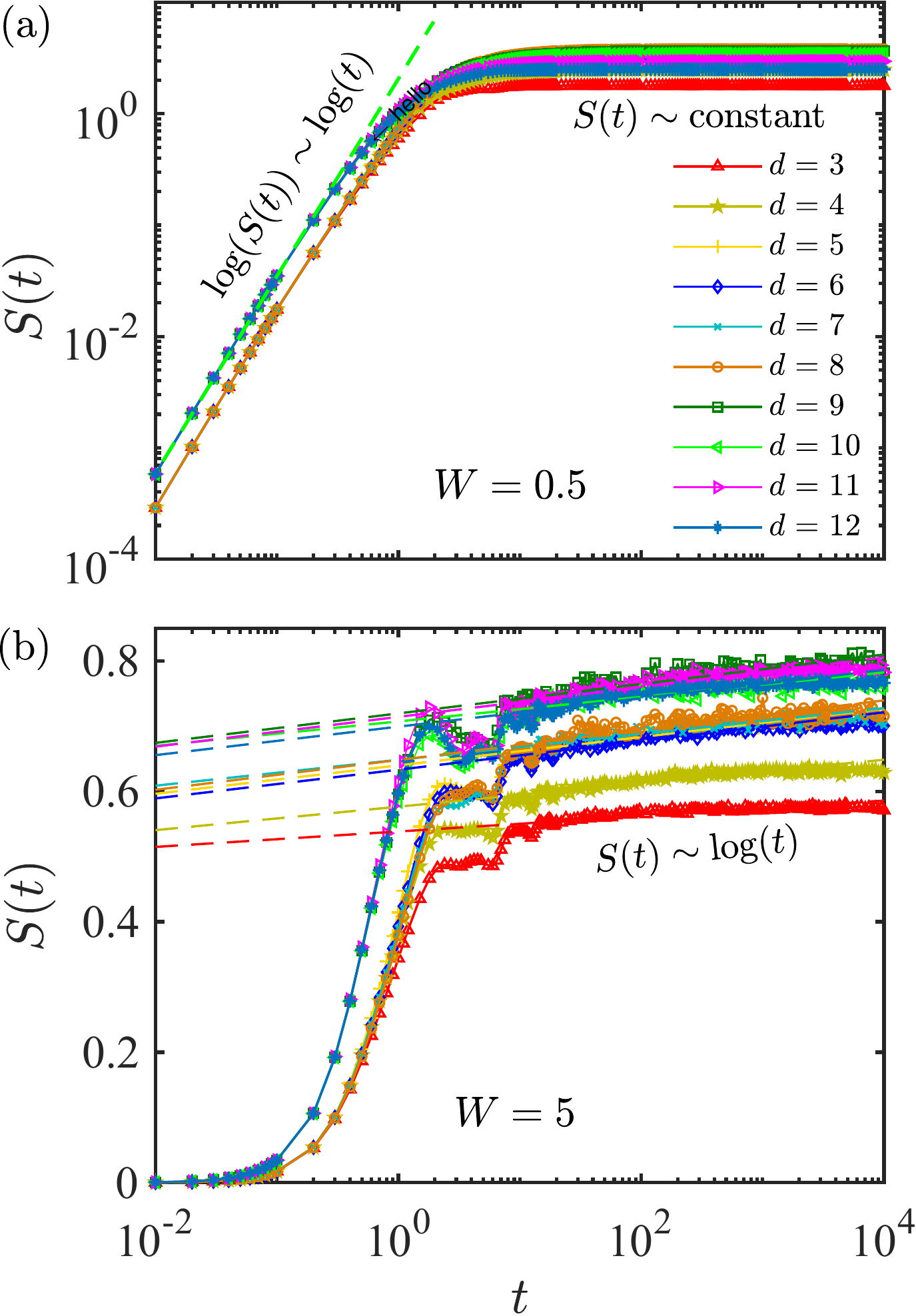}
\caption{{\bf Information scrambling from the dynamics of half-chain EE in the crossover from AL to MBL}. (a) and (b) show the scrambling of EE with $W = 0.5$ (in the ergodic phase) and $W = 5.0$ (in the MBL phase) as a function of $d$. For the ergodic phase we observe the power-law growth of EE $S(t)$ before saturation; while in the MBL phase, we see a logarithmic growth of $S(t)$ in the long time dynamics. All results are 
averaged by 2000 realizations for $L=16$.}
\label{fig-fig4}
\end{figure}

In Fig. \ref{fig-fig3}, we show the $W_c$  and $\nu$ for $d=3-12$; the cases for $d = 1, 2$ are not presented since $W_c = 0$ \cite{supmat}. 
We find that with the increasing of $d$, $W_c$ will saturate to a finite value. This evolution can be described by
\begin{equation}
    W_c(d) = W_c(\infty) + \alpha d^{-\beta}.
    \label{eq-crossoverWc}
\end{equation}
In this fitting with data for $d = 6 - 12$ (dashed line in Fig. \ref{fig-fig3}; see also S1 \cite{supmat}), we obtain $\beta = 4.396$ and $W_c(\infty) = 3.817$. This critical value is slightly greater than $W_c = 3.72$ for $L \le 22$ in Ref. \cite{luitz2015many} and $W_c = 3.8$ for $L \le 24$ in  Ref. \cite{mace2018multifractal,pietracaprina2018shift} coming from the finite size effect (see the open symbols based on 
exact diagonalization for $L\le 14$, 16 and 18; and the scaling of $W_c$ as a function of $L$ in Fig. S4 \cite{supmat}). 
This is an expected feature since in the previous literature, the choose of subspace with $S_\text{z} = 0$ is not compulsory and in principle the same phase transition can be found for
the other $S_\text{z}$. This is in stark contrast to the physics in AL with the uncorrelated disorder, in which $W_c(d) \propto d \ln d$ is expected \cite{anderson1958absence, abrahams1979scaling, 
ueoka2014dimensional, garcia2007dimensional}. 
We also find that $\nu$ are almost unchanged (see inset of Fig. \ref{fig-fig3}), which is also different from the previous literature that $\nu$ depends on the dimension \cite{garcia2007dimensional,tarquini2017critical}. In our model, it is only when $d \ge 6$ that the exponent can satisfy the Harris bound. 
For these reasons, the infinite-range correlated disorder plays an important role in MBL and the associated saturation of $d$ in the thermodynamic limit.

One may naturally ask that why in the standard three dimensional AL, $W_c(3) \simeq 16.5$ and $W_c(d) \propto d \ln d$ in high dimensions \cite{tarquini2017critical}, which is much larger than $W_c$ in our model for MBL. In some models (such as the Aubry-Andr{\' e} model \cite{aubry1980analyticity,iyer2013many}), the correlated disorder is expected to increase the value of $W_c$. To address this issue, we employ the 
transfer matrix to extract the value of $W_c$ (see S4 \cite{supmat}). This method is limited by the finite size effect along the transverse direction. Our 
numerical results for $d = 3 - 5$ are presented by green circles, which are significantly smaller than the value obtained from $r$-statistics. 
In this method, one can see that all the $L^{d-1}$ cross sections are identical, thus will not induce strong scattering between neighboring cross sections, for which reason the physics is 
more approaching the condition with low dimension, hence $W_c$ is small. 

This saturation behavior may also be reflected from the wave function statistics, which are characterized by the participation entropies $S_q$ \cite{luitz2015many,atas2012multifractality},
\begin{equation}
S_q = \frac{1}{1-q} \ln (\sum_{\textbf{n}} |c_{\textbf{n}}|^{2q}),
\end{equation}
via $\psi_d= (c_1,c_2,c_3,\cdots, c_\mathcal{N})^T$ in the virtual lattice. This quality characterizes the localization of the wave function in the virtual lattice with multifractal structure.
In the limit $q\rightarrow 1$, this definition yields the Shannon entropy $S_1=-\sum_{\textbf{n}} |c_{\textbf{n}}|^2 \ln |c_{\textbf{n}}|^2$; while for $q=2$, 
it gives $S_2 = -\ln (\text{IPR})$, with $\text{IPR}=\sum_{\textbf{n}} |c_{\textbf{n}}|^4$ being the inverse participation ratio in the virtual lattice \cite{evers2008anderson,goltsev2012localization}.
In the ergodic ensemble, $c_i$ may satisfy the Gaussian distribution in the ergodic phase \cite{casati1990scaling}: $W(c_i) = (\mathcal{N}/2\pi)^{1/2} \exp(-\mathcal{N} c_i^2/2)$, which 
yields $S_2 = \ln \mathcal{N} - 1.0987$. In the fully localized phase, $c_i$ may be described by the Lorentz distribution \cite{supmat}: $W_c = \gamma/[\pi (c_i^2 + \gamma^2)]$, where $\gamma \sim \delta/\mathcal{N}$, and we have $S_2 = \ln(\frac{3\pi}{2\delta}) + \mathcal{O}(1/\mathcal{N})$. This Lorentz distribution is also consistent with that in the three dimensional AL model \cite{supmat}. 
Between these two regimes, $S_2$ can be well formulated using $S_2 = a_2\ln\mathcal{N}+ b_2$, with $0 \le a_2 \le 1$, thus 
$\delta = \frac{3\pi}{2 e^{b_2} \mathcal{N}^{a_2}}$, which depends weakly on the system size. 
Our results for $d = 3$ and $6$ are shown in Fig. \ref{fig-fig2} (e) and Fig. \ref{fig-fig2} (f). With the increasing of $d$, we find that $a_2$ will changes more and more
slow during the transition from the ergodic phase to the MBL phase. Especially, in the MBL limit, they will approach the same limit (see (g) and (h) in Fig. \ref{fig-fig2}). The saturation of $a_2$ and $b_2$ also indicate a smooth crossover from AL  to MBL. This is expected since the AL will quickly the MBL phase as revealed from Eq. \ref{eq-crossoverWc}. We have also studied the multifractal structure of wave function in S3 \cite{supmat}, which agrees well with the physics in single-particle AL.

{\it MBL from dynamics of EE}. Finally, we explore the smooth crossover from AL to MBL from the dynamics of EE, which measures the information scrambling
with quenched disorder. For two subsystems $A$ and $B$, the reduced density matrix for the state $|\psi\rangle$, is $\rho_A = \text{Tr}_B(|\psi\rangle\langle\psi|)$. 
Then the Von Neumann EE for this partition is given by $S = − \text{Tr}(\rho_A \ln \rho_A)$ \cite{bardarson2012unbounded}. 
To study the dynamics of EE, we choose the N\'eel state as the initial state, and at time $t$, the wave function is obtained through $|\psi(t)\rangle = e^{-i H t} |\psi(0)\rangle$.
In Fig. \ref{fig-fig4}, we show the $\langle S(t)\rangle$ for different dimensions. In the ergodic phase (Fig. \ref{fig-fig4} (a)), we observe a power-law growth of EE in a short time,
which quickly saturate; while in the MBL phase (Fig. \ref{fig-fig4} (b)), we see a clear logarithmic growth of $\langle S(t)\rangle$ in the long time limit 
\cite{bardarson2012unbounded,zhou2017operator}. In Ref. \cite{lukin2019probing}, the long-time logarithmic growth of EE is understood in terms of configuration entropy. As compared 
with the weak disorder condition, the EE is significantly reduced with the strong disorder, indicating that there is no heating in the system and the memory of the initial state is preserved 
during the dynamics \cite{smith2016many,lukin2019probing,brydges2019probing,xu2018emulating}. The small difference in EE for different $d$ indicate that the physics with different $d$ belong to the same universality, thus no phase transition has happened during their crossover.

{\it Conclusions}. In this work, we demonstrate that the MBL can be regarded as an infinite-dimensional AL in the virtual lattice with infinite-range correlated disorder. With the increasing of dimension, we find the critical disorder strength will quickly approach the MBL limit $W_c(\infty) = 3.817$. The infinite-range correlation ensures the saturation of $W_c$ with the increasing of $d$. 
We also demonstrate the smooth crossover from AL to MBL by investigating the dynamics of half-chain EE, 
in which the different dimensions exhibit similar scaling of EE in the same phase. Our results provide a new definition of MBL and give a new theoretical basis for exploring the relationship 
between AL and MBL experimentally. These high dimension physics may be explored using quasiperiodic kicked rotor \cite{PhysRevLett.62.345, PhysRevLett.108.095701}.
Since AL in $d\ge 3$ is well defined \cite{abrahams1979scaling, anderson1958absence}, we expect the phase transition from ergodic phase to the MBL phase is also well defined, 
nevertheless, since localization happens in the virtual lattice, the mobility edge should also be defined in this space, thus it may not be reflected from the real space particle 
diffusion \cite{kohlert2019observation, abanin2019many}.

This work is supported by the National Key Research and Development Program in China (Grants No.
2017YFA0304504 and No. 2017YFA0304103) and the National Natural Science Foundation of China (NSFC) with
No. 11574294 and No. 11774328. M.G. is also supported by the National Youth Thousand Talents Program and the USTC startup funding.

\bibliography{ref}

\clearpage

\renewcommand{\thefigure}{S\arabic{figure}}
\renewcommand{\thesection}{S\arabic{section}}
\renewcommand{\theequation}{S\arabic{equation}}
\renewcommand{\thetable}{S\arabic{table}}
\setcounter{figure}{0}
\setcounter{section}{0}
\setcounter{table}{0}
\setcounter{equation}{0}

\begin{widetext}
\title{XIDIAN UNIVERSITY}
 \maketitle
\appendix
\tableofcontents

\section{S1: Critical disorder strength $W_c$ and the associated critical exponent $\nu$}
\label{sec-S1}

Here we present some more data for the critical disorder strength $W_c$ and the associated critical exponent $\nu$ for different dimensions $d$. In Fig. \ref{fig-figsup1}, we show that the 
level spacing statistics $r$ for $d=2$, which yields $W_c=0$ and $v=2.15$ from the scaling ansatz (Eq. 8 in the main text). 
This means that in the two-dimensional system, any random disorder can lead to the fully localized state. In this case, the mean value of 
$r$ will not reach $\langle r\rangle_{\text{GOE}} = 0.5307$. In Fig. \ref{fig-figsup2}, we show the distribution of $\langle r\rangle$ as a function of the disorder $W$ for dimensions 
$d=4$, 5, 7, 8, 9, 11. In Fig. \ref{fig-figsup2} (a) - (f), we show the transition from the ergodic phase to the MBL phase, where the critical 
disorder strength near the crossing point is marked by gray shaded regimes. Using finite size scaling ansatz, we obtain the 
critical disorder strength $W_c$ and the associated critical exponent $\nu$, showing in Fig. \ref{fig-figsup2} (g) - (l). A complete summary of these values and their variances 
are presented in Table \ref{table-1}. We find that while the critical disorder strengths depend strong on the dimension, the corresponding exponents are shown to be (almost) 
independent of $d$. 

We can extract the critical disorder strength in the MBL phase for $d \rightarrow \infty$ using the following fitting
\begin{equation}
	W_c(d) = W_c(\infty)+\alpha d^{-\beta},
	\label{eq-Wc1}
\end{equation}
for $d = 6 - 12$, which yields $W_c (\infty) = 3.817$, $\alpha = -3794$ and $\beta = 4.396$ (see Fig. \ref{fig-figsup3}). 
In Fig. \ref{fig-figsup4} (a) - (c), we show the MBL transition along $d = L/2$ for $L\leq14$, $L\leq 16$ and $L\leq 18$, respectively. This method
has been used in the previous literature for the searching of $W_c$. We find that the exponent $\nu$ obtained in this way is the same as that
in Fig. \ref{fig-figsup2}. In this fitting, we find that $W_c$ may depend strongly on the value of $L$. Using these three points, together 
with the data from Luitz {\it et. al.} \cite{luitz2015many} for $L\le 22$ and Mac\'e {\it et. al.} \cite{mace2018multifractal,pietracaprina2018shift} for $L \le 24$, 
we can extract the critical disorder strength in the MBL phase using
\begin{equation}
W_c(L) = W_c^\prime(\infty)+\alpha^\prime L^{-\beta^\prime},
	\label{eq-Wc2}
\end{equation}
from which we obtain $W_c^\prime(\infty) = 3.889$, $\alpha^\prime = -58410$ and $\beta^\prime = 4.167$ (see Fig. \ref{fig-figsup4} (d)). The values of 
$\beta$ and $\beta'$ in these two fittings are close to each other. This fitting shows clearly the existence of the saturation point with the increasing of $L$; however, 
notice that since these values are from different sources, the saturation point $W_c'(\infty)$ from the second method 
is slightly different $W_c(\infty)$ from Eq. \ref{eq-Wc1}. Nevertheless, this new fitting and the associated saturation point can still provide valuable 
insight to confirm the existence of the saturation of $W_c$ when $d$ is approaching infinity, which is essential for our major conclusion of the main text. 

\begin{figure}
\centering
\includegraphics[width=0.6\textwidth]{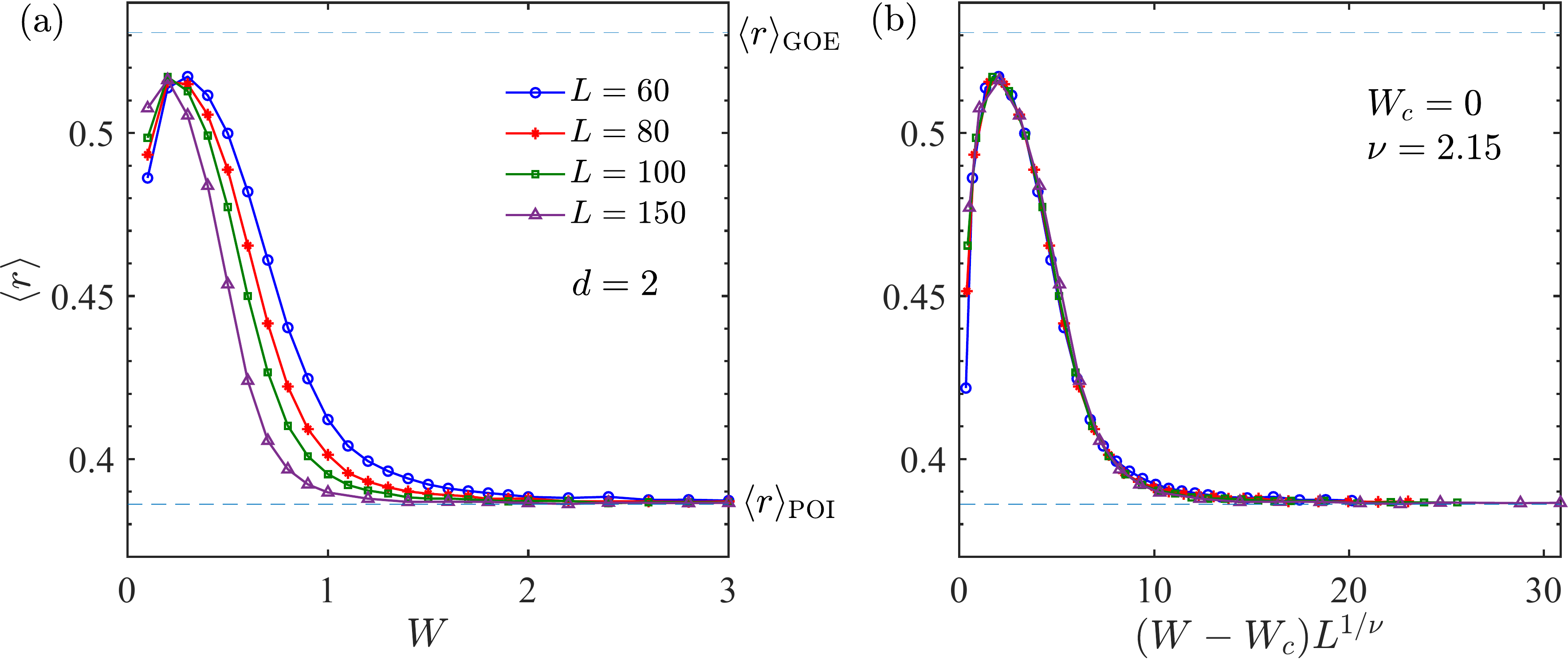}
\caption{{\bf Energy level statistics in two-dimension}. (a) Show the data for $\langle r\rangle$ for different $L$ and (b) show their universal 
	behavior based on Eq. 8 in the main text. We extract $W_c = 0$ and $\nu = 2.15$ by the best fitting. 
	The dashed lines in each figure give $\langle r\rangle_\text{GOE} = 0.5307$ and $\langle r\rangle_\text{PE} = 0.3863$. }
	\label{fig-figsup1}
\end{figure}

\begin{figure}
\centering
\includegraphics[width=0.7\textwidth]{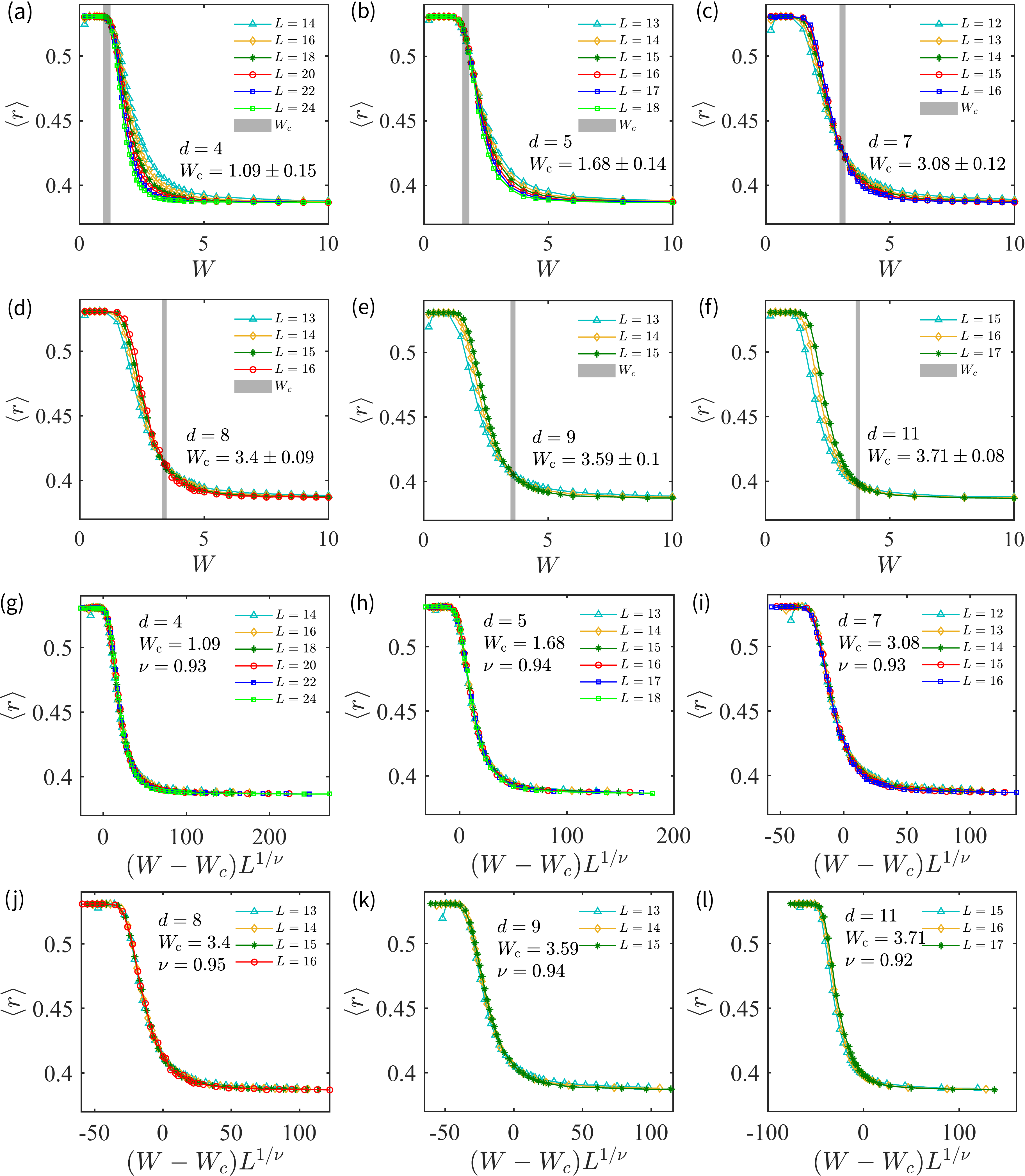}
\caption{\textbf{A summary of energy level statistics and their universal scaling}. Here we consider $d = 4, 5, 7, 8, 11$. In (a) - (f), the crossing points
between different $L$ define the critical boundary between the ergodic phase and the MBL phase (see the gray shaded regimes). In (g) - (l), we show that 
	 all these data can collapse to a single curve based on a single parameter $(W-W_c) L^{1/\nu}$.
	In this way, we obtain $W_c$ and $\nu$, which are shown explicitly in each figure.}
\label{fig-figsup2}
\end{figure}

\begin{table}
\centering
\caption{The critical disorder strength $W_c \pm \delta W$ and the critical exponent $\nu\pm \delta \nu$ for different dimensions $d$ . }
\begin{tabular}{p{1.4cm}<{\centering}p{1.4cm}<{\centering}p{1.4cm}<{\centering}p{1.4cm}<{\centering}p{1.4cm}<{\centering}p{1.4cm}<{\centering}p{1.4cm}<{\centering}p{1.4cm}<{\centering}p{1.4cm}<{\centering}p{1.4cm}<{\centering}p{1.4cm}<{\centering}}    
\toprule 
$d$ & 3 & 4 & 5 & 6 & 7 & 8 & 9 & 10 & 11 & 12 \\ 
\hline 
$W_c$ & 0.726 & 1.094 & 1.682 & 2.38 & 3.08 & 3.403 & 3.59 & 3.675 & 3.71 & 3.74 \\
\hline 
$\delta W$ & 0.17 & 0.15 & 0.142 & 0.106 & 0.116 & 0.092 & 0.095 & 0.086 & 0.08 & 0.07\\
\hline
$\nu$ & 0.956 & 0.929 & 0.937 & 0.91 & 0.934 & 0.948 & 0.94 & 0.935 & 0.90 & 0.91\\
\hline
$\delta \nu$ & 0.04 & 0.049 & 0.044 & 0.042 & 0.046 & 0.045 & 0.038 & 0.048 & 0.041 & 0.037 \\
\toprule
\end{tabular}
\label{table-1}
\end{table} 

\begin{figure}
\centering
\includegraphics[width=0.5\textwidth]{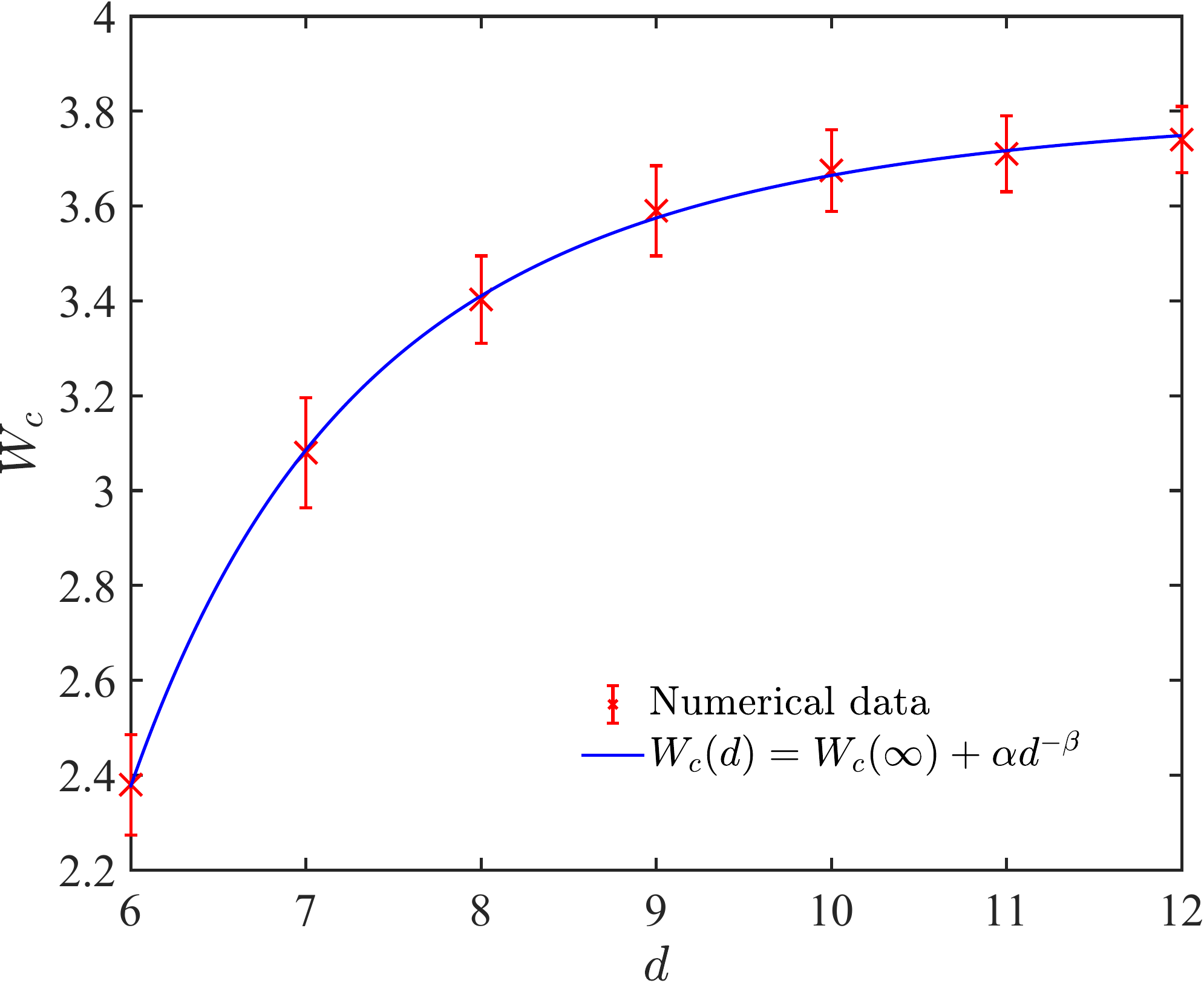}
\caption{{\bf Scaling of the critical disorder strength}. The red cross with error bars are the critical disorder strength for different dimension $d$
	from level spacing statistics. The solid line is our best fitting using Eq. \ref{eq-Wc1}, which yields $W_c (\infty) = 3.817$, $\alpha = -3794$ and $\beta = 4.396$. 
	This data is re-plotted in Fig. 3 in the main text as a function of $1/d$ with the dashed line.}
\label{fig-figsup3}
\end{figure}

\begin{figure}
\centering
\includegraphics[width=0.6\textwidth]{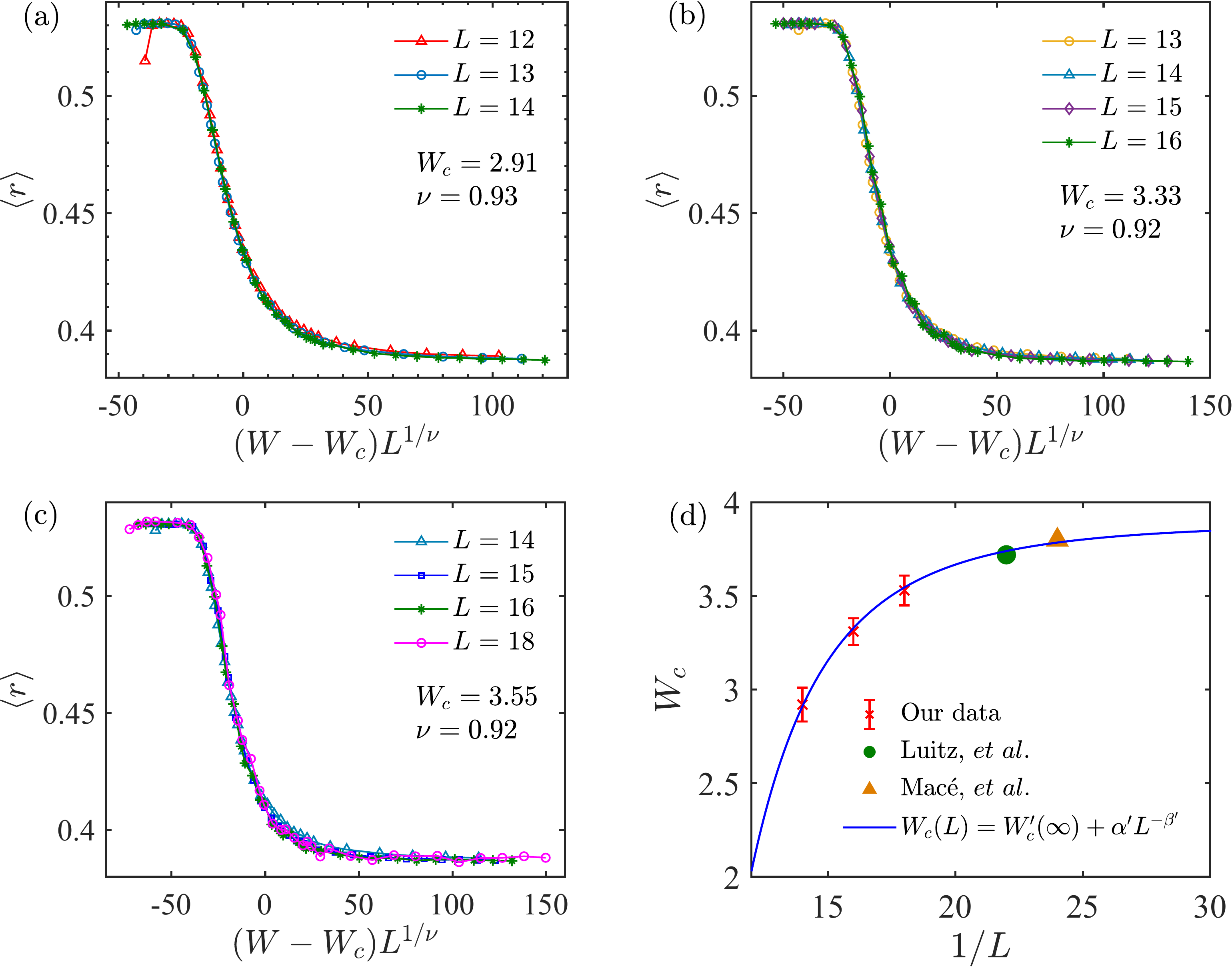}
\caption{{\bf MBL transition along the line $d = L/2$}. (a) - (b) show the universal scaling of $\langle r\rangle$ for different $L$ using Eq. 8 in the main text. 
	(d) The red cross with error bars are from (a) - (c) and the filled symbols are given by Luitz {\it et. al.} for $L \le 22$ ($W_c=3.72$) \cite{luitz2015many} 
	and Macé {\it et. al.} for $L \le 24$ ($W_c=3.8$) \cite{mace2018multifractal,pietracaprina2018shift}. The blue solid line is the best for these five points
	using Eq. \ref{eq-Wc2}, which yields $W_c^\prime(\infty) = 3.889$, $\alpha^\prime = -58410$ and $\beta^\prime = 4.167$. }
\label{fig-figsup4}
\end{figure}

\section{S2: Analytical results for participation entropies $S_q$ in the weak and strong disorder limits}
\label{sec-S2}

In the ergodic phase, $c_i$ in the wave function $\psi_d = (c_1,\cdots,c_{\mathcal{N}})^T$ in GOE satisfies the 
Gaussian distribution (see Ref. \cite{casati1990scaling})
\begin{equation}
W_{\text{GOE}}(c_i) = (A/\sigma\pi)^{1/2} \exp(-A c_i^2/\sigma).
\label{eq-WGOE}
\end{equation}  
In Fig. \ref{fig-figsup5} (a) - (d), we present our numerical results for some of these wave function components. We see that the distribution of $c_i$ in 
the ergodic phase satisfies $W_{\text{GOE}}(c_i)$, with $A\approx \mathcal{N}$ and $\sigma\approx 2$. However, in the MBL phase, we find the wave function components 
satisfy the Lorentz distribution
\begin{equation}
W_{\text{MBL}}(c_i) = \frac{1}{\pi}\frac{\gamma}{x^2+\gamma^2}, \quad \gamma=\frac{\delta}{\mathcal{N}},
\label{eq-WMBL}
\end{equation}  
where $\delta$ depends weakly on $W$ and $\mathcal{N}$ in some way. Some examples are shown in Fig. \ref{fig-figsup5} (e) - (h). Let us assume
\begin{equation}
S_q = \frac{1}{1-q} \ln (\sum_{\textbf{n}}^{\mathcal{N}} |c_{\textbf{n}}|^{2q}) = \frac{1}{1-q} \ln ( \left\langle \sum_{\textbf{n}}^{\mathcal{N}} |c_{\textbf{n}}|^{2q} \right\rangle \mathcal{N}),
\end{equation}
then in the ergodic phase, we have 
\begin{equation}
\left\langle \sum_{\textbf{n}}^{\mathcal{N}} |c_{\textbf{n}}|^{2q} \right\rangle = \int_{-1}^{1} C W_{\text{GOE}}(x) x^{2q} \text{d} x = \frac{2^q [\Gamma (q+\frac{1}{2}) +\Gamma(q+\frac{1}{2},\frac{\mathcal{N}}{2})]}{\sqrt{\pi}\mathcal{N}^{q}\text{Erf}(\sqrt{\mathcal{N}/2})},
\end{equation}
where $C=\text{Erf}(\sqrt{\mathcal{N}/2})$ is the normalized constant for $|c_i|\leq 1$, 
$\text{Erf}(x)$ is the Gauss error function, $\Gamma(\cdot)$ and $\Gamma(\cdot, \cdot)$ are the Gamma and incomplete Gamma functions, respectively. 
So for any positive integer $q$ and large $\mathcal{N}$, we have
\begin{equation}
S_q \simeq \ln\mathcal{N} + b_q, \quad b_q = \frac{\ln[2^q\Gamma(q+1/2)/\sqrt{\pi}]}{1-q}.
\label{eq-SqGOE}
\end{equation}
which means $a_q=1$. We find that Eq. \ref{eq-SqGOE} is consistent with the numerical results (see Fig. \ref{fig-figsup6}).
When $q = 0$, $\Gamma(1/2) = \sqrt{\pi}$, thus $b_0 = 0$, which is shown in Fig. \ref{fig-figsup7} (a). 
In the MBL phase with the Lorentz distribution, we find 
\begin{equation}
\left\langle \sum_{\textbf{n}}^{\mathcal{N}} |c_{\textbf{n}}|^{2q} \right\rangle = \int_{-1}^{1} D W_{\text{MBL}}(x) x^{2q} \text{d} x = \frac{_2F_1 (1,\frac{1}{2}+q,\frac{3}{2}+q,-\frac{1}{\gamma^2})}{(1+2q)\gamma \arctan (\frac{1}{\gamma})},
\end{equation}
where $D=\frac{2}{\pi} \arctan(\frac{1}{\gamma})$ is the normalized constant for $|c_i|\leq 1$ and $_2F_1$ is the hypergeometric function. So, we find
\begin{equation}
S_q = \frac{1}{1-q} \ln[\frac{_2F_1(1,1/2+q,3/2+q,-1/\gamma^2)\mathcal{N}}{(1+2q)\gamma \arctan(1/\gamma)}].
\label{eq-SqMBL}
\end{equation}
For $q=2$ and large $\mathcal{N}$, we have
\begin{equation}
S_2 =\ln(\frac{3\pi}{2\delta}) + \mathcal{O}(\frac{1}{\mathcal{N}}).
\label{eq-S2MBL}
\end{equation}
For $\delta\in[1, 2]$, the result given by Eq. \ref{eq-S2MBL} is consistent with the numerical results in Fig. \ref{fig-figsup6}. So combined with the assumption 
$S_2 = a_2\ln \mathcal{N} + b_2$, we have
\begin{equation}
\delta = \frac{3\pi}{2 e^{b_2} \mathcal{N}^{a_2}}, \quad \gamma = \frac{3\pi}{2 e^{b_2} \mathcal{N}^{1+a_2}}
\end{equation}
This relation means that for $0 \le a_2 \le 1$, the width of the Lorentz distribution will decrease slightly faster than $1/\mathcal{N}$. These two coefficients 
depend strongly on the value of 
disorder strength (see Fig. 2 in the main text). This relation will be useful for us to understand the value of $\delta$ in Fig. \ref{fig-figsup5}, 
	 in which for $a_2 = 0.1938$ and $b_2 = -0.1201$ 
(the data for $d=10$ and $W=5$ from Fig. 2 (g) - (h) in the main text), we have $\delta = 1.097$, in consistent with our estimation $\delta \sim 0.945 - 1.64$. Here the 
tiny difference comes from the fluctuation of wave function components in the MBL phases. It is necessary to emphasize that the Gaussian distribution of $c_i$ in
the ergodic phase and the Lorentz distribution of $c_i$ in the localized phase regime for the results in Fig. \ref{fig-figsup5} can be well reproduced by the three
dimensional Anderson localization model.

\begin{figure}
	\centering
	\includegraphics[width=0.8\textwidth]{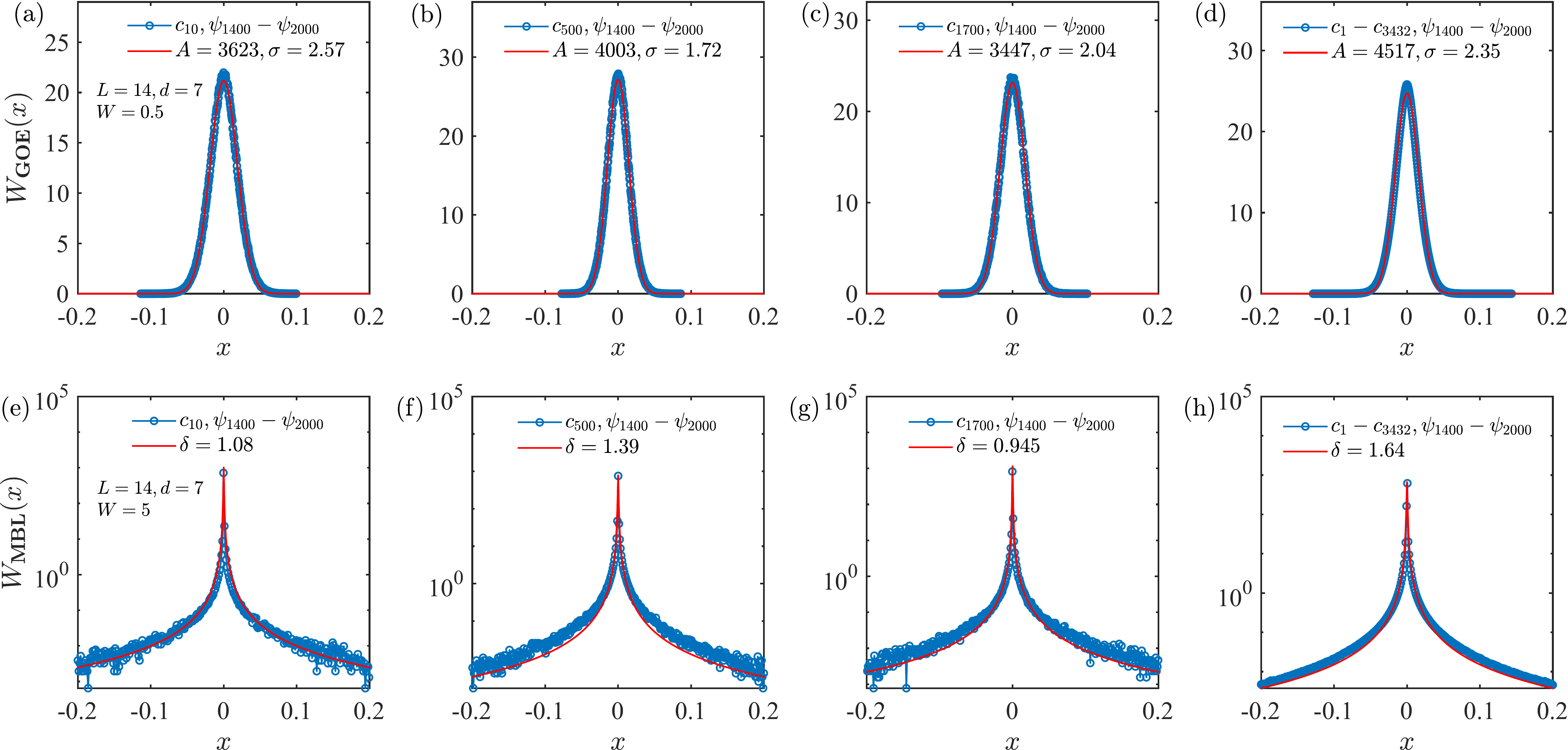}
	\caption{{\bf Statistics of wave function components}. The upper panel and down panel have used $W = 0.5$ and $W = 5$, respectively, for $L = 14$, $d = 7$, thus $\mathcal{N}= 3432$. 
	In the first, second and third columns, we consider the statistics for $c_{10}$, $c_{500}$, $c_{1700}$ for eigenstate $\psi_{1400}$ to $\psi_{2000}$ (in the middle of the spectra). 
	The last column shows the statistics for all wave function components from $c_1$ to $c_{3432}$. In the upper panel, the solid curve is fitted using the Gaussian distribution (Eq. \ref{eq-WGOE}) with width $\sigma/
	\mathcal{N}$, and in the down panel, it is fitted using the Lorentz distribution (Eq. \ref{eq-WMBL}) with the width by  $\delta/\mathcal{N}$.}
	\label{fig-figsup5}
\end{figure}

\begin{figure}
	\centering
	\includegraphics[width=0.5\textwidth]{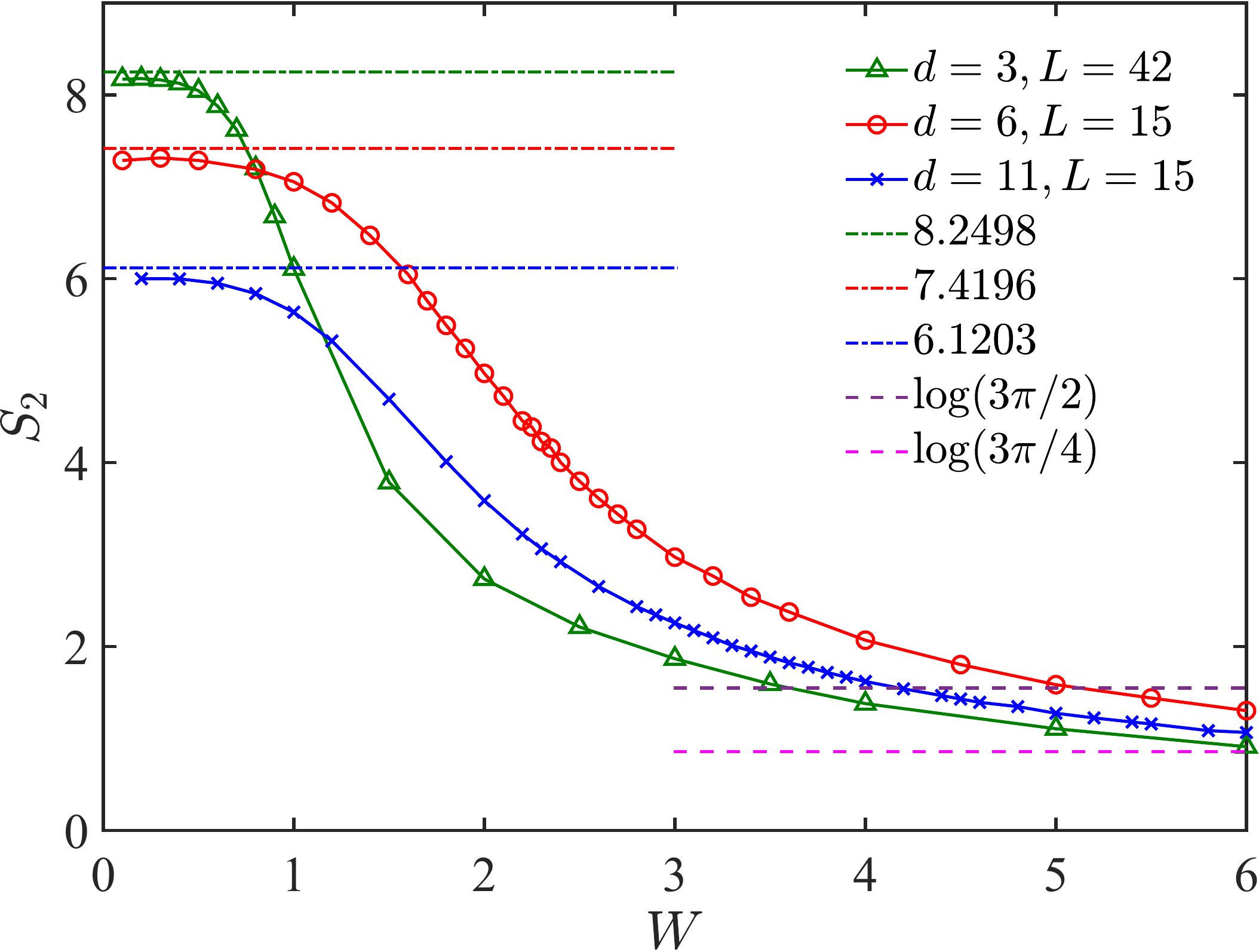}
	\caption{{\bf Numerical and analytical value of $S_2$}. The symbols represent numerical results and the horizontal dashed lines are given by Eq. \ref{eq-SqGOE} for ergodic phase with $\sigma =1$, 
	in which $\mathcal{N} = C_{42}^3$, $C_{15}^{6}$ and $C_{15}^{11}$ will give $S_2$ = 8.2498, 7.4196 and 6.1203, respectively; and Eq. \ref{eq-S2MBL} for MBL phase with $\delta = 1$ 
	(with $S_2 = \ln(3\pi/2)$) and $\delta = 2$ (with $S_2 = \ln(3\pi/4)$).}
	\label{fig-figsup6}
\end{figure}

\begin{figure}
\centering
\includegraphics[width=0.7\textwidth]{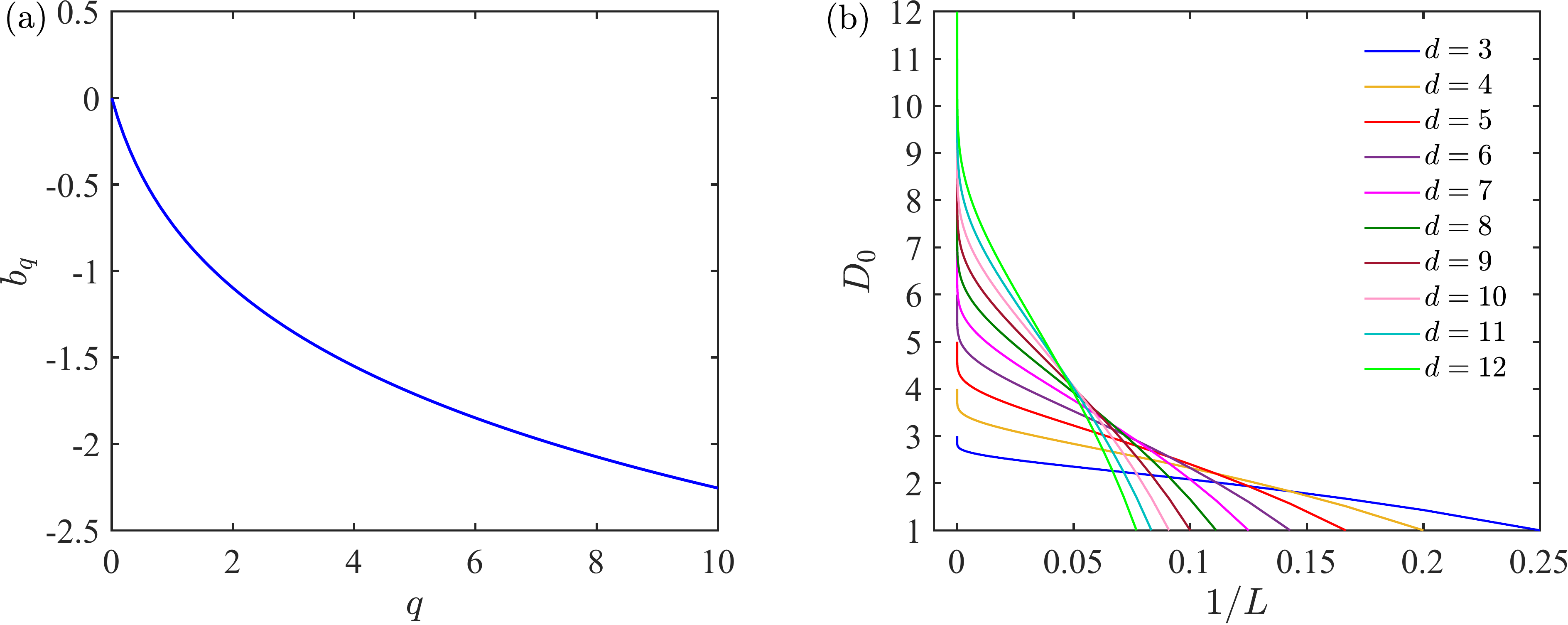}
\caption{{\bf Constant term $b_q$ in $S_q$ and the dimension of the system from $D_0$}. (a) $b_q$ as a function of $q$ given by Eq. \ref{eq-SqGOE}. (b) $D_0$ as a function of $1/L$ for $d = 3 - 12$.  
	In the thermodynamic limit $L\rightarrow \infty$, $D_0=d$ (Eq. \ref{eq-d0}), which means the negligible of boundary effect.}
	\label{fig-figsup7}
\end{figure}

\section{S3: Multifractal of the wave function in the virtual lattice across the critical boundary and features identical to single particle AL}
\label{sec-S3}

It has been widely explored that during the Anderson transition from the extended phase to the localized phase, 
the change of wave function can be reflected from its multifractal structures. In this section, we analysis 
the multifractal structure of $\psi_i$ in the virtual lattice. We can define the $q$-th moment as \cite{vasquez2008multifractal}
\begin{equation}
P_q(L) = \sum_i |\psi_i|^{2q} \propto L^{-\tau(q)}.
\end{equation} 
The mass exponent $\tau(q)$ is defined as
\begin{equation}
\tau(q) = - \lim_{L\rightarrow \infty} \frac{\ln \langle P_q(L) \rangle}{\ln L} \equiv d(q-1) + \Delta_q \equiv D_q(q-1),
\end{equation} 
where $\Delta_q$ is the anomalous dimension with $\Delta_0=\Delta_1=0$ and $D_q$ is the generalized fractal dimension. The similarity dimension $D_0$ equals 
to the Euclidean dimension, given by
\begin{equation}
D_0 = -\tau(0) = \frac{\ln\langle P_0(L) \rangle}{\ln L} = \frac{\ln \sum_i^{\mathcal{N}} |\psi_i|^0}{\ln L} = \frac{\ln \mathcal{N}}{\ln L} = \frac{\ln C_L^d}{\ln L} = \frac{\ln \frac{L!}{d!(L-d)!}}{\ln L}.
\end{equation}
For $L\gg d$, we find that 
\begin{equation}
D_0 \simeq \frac{\ln [L(L-1)\cdots (L-(d-1))]}{\ln L} \simeq \frac{\ln [ L^d (1-\frac{d(d-1)}{2L})]}{\ln L} \simeq d.
\label{eq-d0}
\end{equation}
So, in the thermodynamic limit, we have $D_0 = d$ (see \ref{fig-figsup7} (b)). This is slightly different from the $d$-dimensional Anderson model with $D_0 = \frac{\ln L^d}{\ln L} \equiv d$, which is 
independents of size $L$. This feature reflects the finite boundary in our virtual lattice (see Fig. 1 (b) and Fig. 1 (c) in the main text), in which the boundary effect is negligible when $L$ 
is large enough. The singularity spectrum $f(\alpha)$ and mass exponent $\tau(q)$ are related to each other via the Legendre transformation,
\begin{equation}
f(\alpha_q) = \alpha_q q -\tau(q),  \quad \alpha_q = \frac{d\tau(q)}{dq}
\end{equation}
 Numerically, we can evaluate $f(\alpha_q)$ and $\alpha_q$ by \cite{vasquez2008multifractal}
\begin{equation}
\alpha_{q}=-\lim_{L \to \infty} \frac{1}{\ln L}\left\langle\sum_{k=1}^{\mathcal{N}} \delta_{k}(q, L) \ln \delta_{k}(1, L)\right\rangle, \quad 
f(\alpha_q)= -\lim_{L \to \infty} \frac{1}{\ln L}\left\langle\sum_{k=1}^{\mathcal{N}} \delta_{k}(q, L) \ln \delta_{k}(q, L)\right\rangle,
\end{equation}
where $\delta_{k}(q, L) \equiv |\psi_i|^{2q} /P_q(L) $. In Fig. \ref{fig-figsup8}, we show the mass exponent $\tau_q$, generalized fractal dimension $D_q$ and singularity spectrum $f(\alpha_q)$. The maximum value of $f(\alpha)/d$ is less than 1 due to the finite size effect (see Fig. \ref{fig-figsup8} (d) and (h)), whereas in the AL model, this value is equal to 1. 
These features mean that the wave functions in the MBL phase have the same multifractal structure as that for the single-particle AL models. 
Noticed that in our model, the localization happens in the virtual lattice, instead of the real space. 

\begin{figure}
\centering
\includegraphics[width=0.8\textwidth]{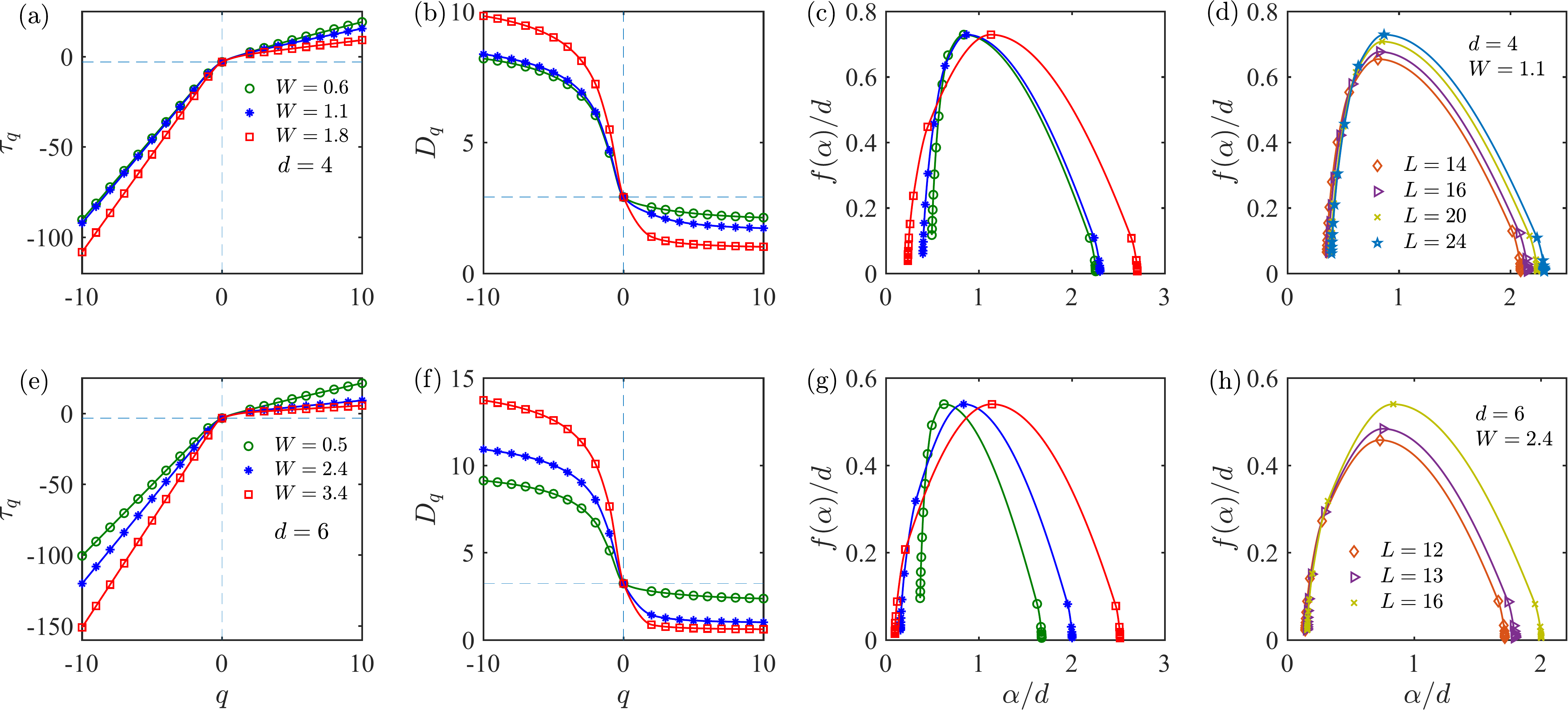}
\caption{ \textbf{Multifractal analysis of wave function and the similarity to single-particle AL}. 
	The first and second columns give $\tau_q$, $D_q$ as a function of $q$, the third column give $f(\alpha)/d$ as a function of 
	$\alpha/d$ and the last column give $f(\alpha)/d$ as a function of $L$. The upper panel (a) - (c) give data for $L = 24$ and $d = 4$ ($W_c = 1.09$) and the down panel (e) - (g) for $L=16$ and $d = 6$ ($W_c = 2.38$).}
\label{fig-figsup8}
\end{figure}

\section{S4: Transfer matrix methods in high dimensional AL with correlated disorder and the estimation of $W_c$}
\label{sec-S4}

In the $d$-dimensional AL model, the critical disorder strength is shown to be proportional to $d$ (some more numerical results in Ref. \cite{tarquini2017critical} 
show that $W_c \propto d\ln(d)$). For example, when $d = 3$, $W_c = 7.9$; when $d = 4$, $W_c = 17.0$ and when $d = 5$, $W_c = 28.4$. This critical boundary is much larger 
than the critical strength obtained in Fig. 3 in the main text. To understand this difference, we try to determine the $W_c$ using the transfer method. The drawback of this 
method is that it can only deal with some physics in small quantum systems. We treat the $d$ dimensional model with the correlated disorder and with the uncorrelated disorder 
in equal footing, so as to provide useful insight into the physics in our model.

To this end, we rewrite the Hamiltonian in the $d$-dimensional virtual lattice as
\begin{equation}
H_d =  \sum_{\langle\textbf{n},\textbf{n}^\prime\rangle} \frac{J}{2} (\hat{c}_\textbf{n}^\dagger \hat{c}_{\textbf{n}^\prime} + h.c.) 
	 + \sum_{\textbf{n}} \xi_{\textbf{n}}  \hat{c}_{\textbf{n}}^\dagger \hat{c}_{\textbf{n}},
	\label{eq-Hd}
\end{equation}
where we ignore the boundary condition, thus $n_i\in[1,L]$ ($i=1,\cdots,d$). For $d=1$, Eq. \ref{eq-Hd} can be rewritten as
 \begin{equation}
H_1 =  \sum_{n} \frac{J}{2} (\hat{c}_{n}^\dagger \hat{c}_{n+1} + h.c.) 
	 + \sum_{n} h_{n}  \hat{c}_{n}^\dagger \hat{c}_{n}.
 	 \label{eq-H1}
\end{equation}
Then we have $c_{n+1} + c_{n-1} = \frac{2}{J}(E-h_n) c_n$, where $c_n$ is the wave function on the 
$n$-th site and $E$ is the eigenvalue. This relation can be written in the transfer matrix form as 
\cite{hoffmann2012computational}
\begin{equation}
\left(\begin{array}{c}{c_{n+1}} \\ {c_{n}}\end{array}\right) =\left(
\begin{array}{cc}
{ 2(E-h_n)} & {-1} \\ {1} & {0}
\end{array}
\right) \left(\begin{array}{c}{c_{n}} \\ {c_{n-1}}\end{array}\right)\equiv T_n \left(\begin{array}{c}{c_{n}} \\ {c_{n-1}}\end{array}\right),
\end{equation}
where $T_n$ is the transfer matrix, and we have set $J = 1$. In general, the transfer matrix is given by 
\begin{equation}
\left(\begin{array}{c}{\mathbf{c}_{n+1}} \\ {\mathbf{c}_n}\end{array}\right) =\left(
\begin{array}{cc}
{\mathcal{H}_n } & {-I} \\ {I} & {0}
\end{array}
\right) \left(\begin{array}{c}{\mathbf{c}_n} \\ {\mathbf{c}_{n-1}}\end{array}\right) \equiv T_n \left(\begin{array}{c}{\mathbf{c}_n} \\ {\mathbf{c}_{n-1}}\end{array}\right),
\label{eq-TM}
\end{equation}
where $ \mathbf{c}_n $ is the wave function on the $n$-th cross-section (the dimension is $d-1$ with size $L^{d-1}$) and $\mathcal{H}_n$ is the Hamiltonian of the $n$-th 
cross-section. Eq. \ref{eq-TM} may be solved recursively for arbitrary initial conditions $\mathbf{c}_1$ and $\mathbf{c}_0$, then, the wavefunction at $N+1$, $N$ are 
given by
\begin{equation}
\left(\begin{array}{c}{\mathbf{c}_{n+1}} \\ {\mathbf{c}_n}\end{array}\right) = T_NT_{N-1}\cdots T_1 \left(\begin{array}{c}{\mathbf{c}_1} \\ {\mathbf{c}_0}\end{array}\right)
\end{equation}
The propagation of the transfer matrix is shown in Fig. \ref{fig-figsup9}. To get the amplitudes, we can calculate the product of the transfer matrix,
\begin{equation}
\mathcal{T}_N = \prod_{n=1}^N T_n = T_N\cdots T_2Q_1R_1 = T_N\cdots T_3Q_2R_2R_1 = Q_NR_N\cdots R_1,
\end{equation}
where we use the QR decomposition ($T \equiv QR$). In the thermodynamic limit, we can define
\begin{equation}
\Lambda  = \lim_{N\rightarrow\infty} \ln(\mathcal{T}_N^\dagger \mathcal{T}_N)^{1/2N} = \lim_{N\rightarrow\infty} \ln (R_1^\dagger\cdots R_N^\dagger R_N\cdots R_1)^{1/2N}  = \lim_{N\rightarrow\infty} \ln ( \prod_n^N R_n^2 )
\end{equation}
with eigenvalues $\lambda_i$ ($i=1,\cdots,L$), where $\lambda_i$ are the Lyapunov exponents. Then, the localization length $\xi$ is given by  $\xi = (\min_{i=1\cdots L}{|\lambda_i|})^{-1}$. From the 
theorem of Oseledec \cite{hoffmann2012computational}, we assume $\Lambda$ exist. 

In Fig. \ref{fig-figsup10} (a) - (b), we show the transfer matrix results of Anderson model 
\begin{equation}
	H_\text{AL} = \sum_{\langle i,j\rangle} (\hat{c}_i^\dagger \hat{c}_j + h.c.) + \sum_i \varepsilon_i \hat{c}_i^\dagger \hat{c}_i,
\label{eq-HA}
\end{equation} 
where $\varepsilon_i \in [-W,W]$ is the on-site random potential. The on-site correlation is $\langle\varepsilon_i^2\rangle = \frac{W^2}{3}$. However, the on-site correlation for Eq. \ref{eq-Hd} is given by $\langle\xi_{\textbf{n}}^2\rangle = \frac{dW^2}{3}$. For $d=3$ ($d=5$), we get $W_c = 7.9\pm 0.45$ ($W_c = 28.4\pm0.45$), which is consistent with the previous literature $W_c \simeq 8.175$ ($W_c\simeq 28.75$) \cite{tarquini2017critical}. 

In Fig. \ref{fig-figsup10} (c) - (d), we show the localization length $\xi(L)/L$ as a function of $W$ for Eq. \ref{eq-Hd}. The critical disorder $W_c\simeq 0.03$ and $W_c\simeq 0.28$ for $d=3$ and $d=5$. 
One can see that the critical disorder $W_c$ for the Anderson model with the correlated disorder is much smaller than the Anderson model without correlated disorder. 
Noticed that this value is still not converged due to the large localization length; however, its small value as compared with Fig. \ref{fig-figsup10} (a) - (b), can still reflect some unique features
of our idea for MBL. This result can be understood from the fact that in Fig. \ref{fig-figsup9}, all the cross-sections have the same energy levels, thus the disorder will not induce strong scattering between the different states in the neighboring cross-sections. For this reason, although the physics is essentially a high dimensional model, the scattering is more likely to be a low dimensional one (see Fig. \ref{fig-figsup9} (b)), thus $W_c$ can be small. 

\begin{figure}
\centering
\includegraphics[width=0.55\textwidth]{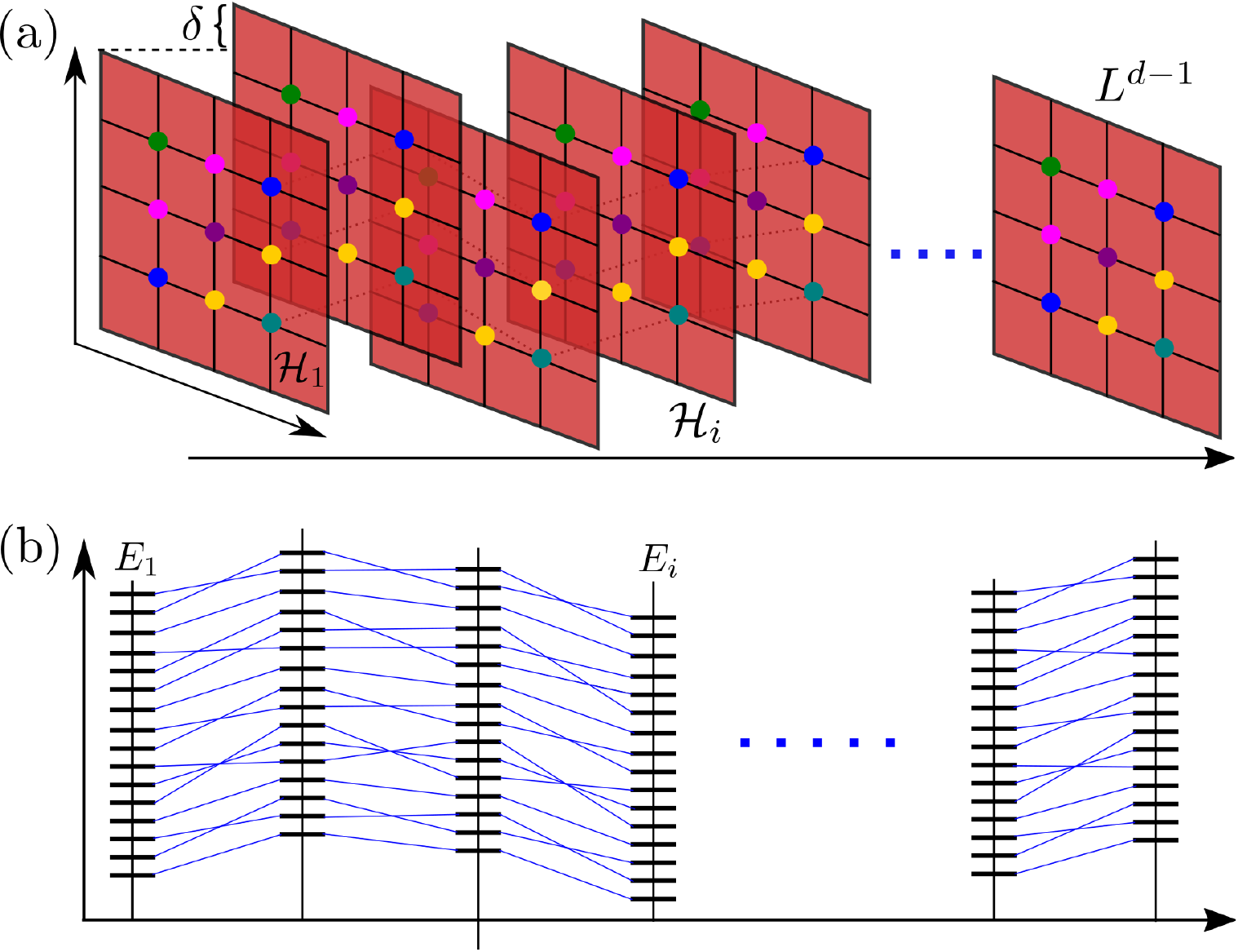}
\caption{  \textbf{Propagation of the transfer matrix.} (a) Schematic diagram of the quasi-one-dimensional  chain along the x-direction of cross-section $L^{d-1}$. $\mathcal{H}_{i}$ is the Hamiltonian in $i$-th cross-section. The sites at the same position in different cross-sections only differ by a global random potential. For this reason, although the model itself is $d$ dimension, its effect is more likely to be a low dimensional one. (b) Scattering between different states between neighboring cross-sections, which have identical scattering matrix when each cross-section is diagonalized. }
\label{fig-figsup9}
\end{figure} 

\begin{figure}
\centering
\includegraphics[width=0.55\textwidth]{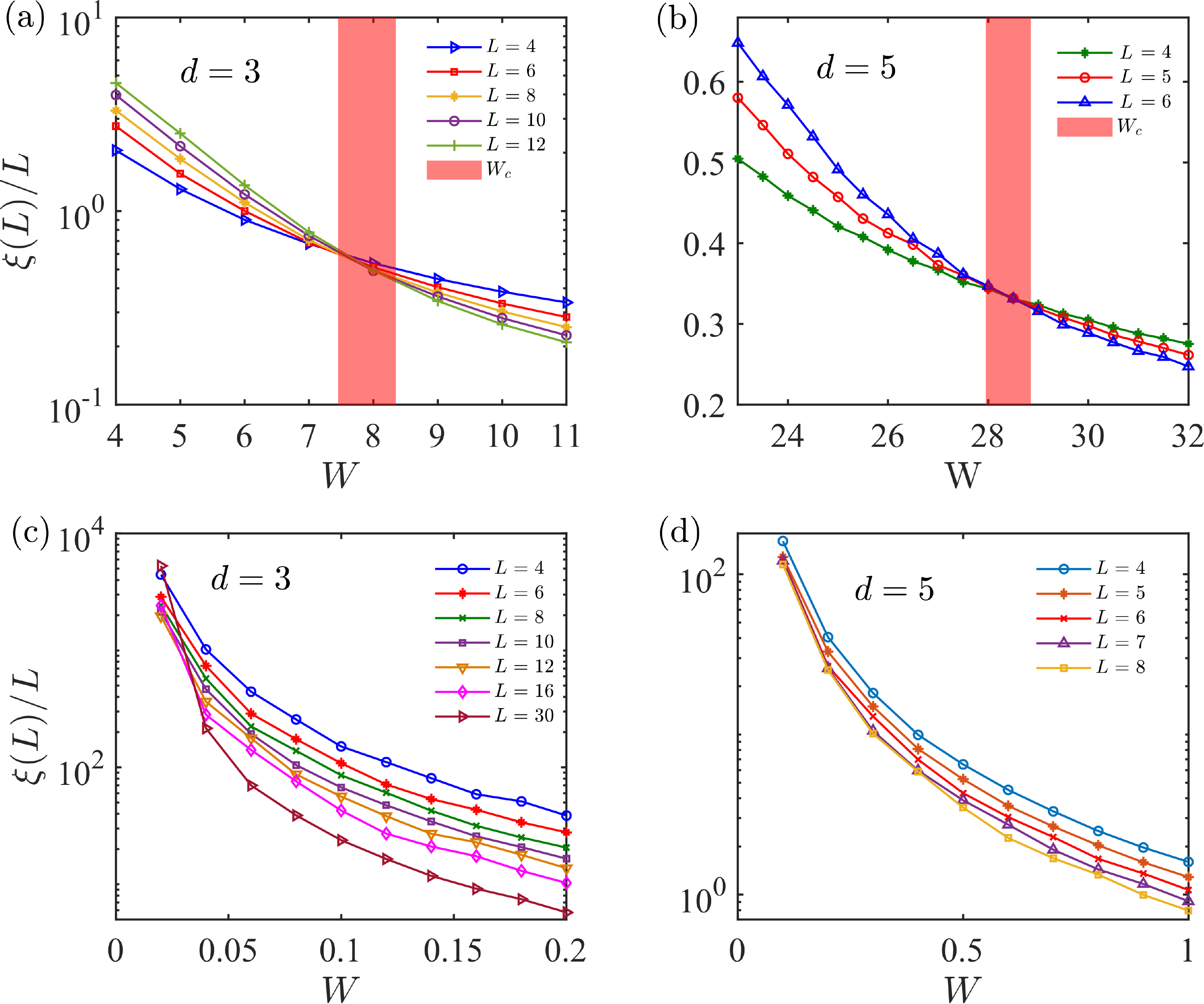}
\caption{  \textbf{Transfer matrix results for the $d$-dimensional AL without correlation (upper panel) and with correlation (down panel)}. Here we plot $\xi(L)/L$ as a function of the disorder $W$. 
	(a) and (b) Anderson model in Eq. \ref{eq-HA} for $d=3$ and $d=5$, and the critical disorder are $W_c = 7.9\pm 0.45$ and $W_c = 28.4\pm0.45$ (see the red shaded regimes), respectively. 
	These critical strengths are consistent with $W_c=8.175$ for $d=3$ and $W_c=28.75$ for $d=5$ presented in Ref. \cite{tarquini2017critical} with much larger system sizes.  
	For comparison, in (c) and (d) we present the AL with correlated disorder for Eq. \ref{eq-Hd} for $d=3$ and $d=5$, which yield $W_c = 0.03\pm0.01$ and $W_c = 0.28 \pm0.08$, respectively. Discussion about this result is presented in the text.}
\label{fig-figsup10}
\end{figure}

\section{S5: MBL with fermion and boson models, physics with random interaction and the general picture for MBL with $\mathbb{Z}_2$ symmetry}
\label{sec-S5}

This section address three fundamental issues. Firstly, how to apply our new definition of MBL to fermion and boson models; Secondly, what will 
happen in the presence of many-body random interaction, instead of random chemical potential; and finally, how to understand the physics with 
$\mathbb{Z}_2$ symmetry, for example in the XYZ model. We aim to show that in all these different cases, our basic idea that MBL can be regarded
as an infinite-dimensional AL with infinite-range correlated disorder is always held. 

\subsection{A: Fermi-Hubbard model with the conserved number of particle} 

The same picture for fermion and boson can be realized in the following way. Let us consider the spinless Fermi-Hubbard model in the following way. For the model \cite{oganesyan2007localization},
\begin{equation}
	H = -t \sum_{i} \hat{c}_{i}^\dagger \hat{c}_{i+1} + \text{H.c.} + U \hat{n}_i \hat{n}_{i+1} + v_i \hat{n}_i,
\end{equation}
where $U$ is the interaction between the neighboring sites and $v_i$ is the random potential. We assume $d$ particles in this chain, thus we can define the following
basis
\begin{equation}
	|\phi_{\bf n} \rangle = |n_1 n_2 n_3 \cdots n_d\rangle = \hat{c}_{n_1}^\dagger \hat{c}_{n_2}^\dagger \hat{c}_{n_3}^\dagger \cdots \hat{c}_{n_d}^\dagger |0\rangle, \quad 1\le n_i < n_{i+1} \le L, 
	\label{eq-phin}
\end{equation}
which means that only $n_i$-th site is occupied by one fermion particle. We find that this picture is the same as that for the spin model, thus is not repeated here. 

In the following, we discuss how to incorporate this idea for fermions with spin degree of freedom. To this end, let us introduce the concept of flavor $s$, which accounts 
for the possible the spin degree of freedom. Then the model may be written as \cite{luitz2017ergodic}
\begin{equation}
	H = -t \sum_{i,s,s'} \hat{c}_{i,s}^\dagger \hat{c}_{i+1,s} + \text{H.c.} + U_{s,s'} \hat{n}_{i,s} \hat{n}_{i+1,s'} + v_{i,s} \hat{n}_{i,s},
\end{equation}
where $\hat{n}_{i,s} = \hat{c}_{i,s}^\dagger \hat{c}_{i,s}$. Let us consider the simplest case, as shown in Fig. \ref{fig-figsup11}, in which the spin degree of freedom can
be denoted as some kind of flavor (spin configuration) in each site. In this case, the basis can still be defined in a similar way as Eq. \ref{eq-phin}, with
\begin{equation}
	|\phi_{\bf n} \rangle = |n_{1s_1} n_{2s_2} n_{3s_3} \cdots n_{ds_d}\rangle = \hat{c}_{n_{1s_1}}^\dagger \hat{c}_{n_{2s_2}}^\dagger \hat{c}_{n_{3_s3}}^\dagger \cdots \hat{c}_{n_{ds_d}}^\dagger |0\rangle,
	\label{eq-SpinFermi}
\end{equation}
where $n_{is_i}$ represents the creation of a fermion with spin $s_i$ at the $n_i$-th site and the boundary condition is 
$1\leq n_{1s_1}\leq n_{2s_2}\leq n_{3s_3}\leq \cdots \leq n_{ds_d}\leq L$ ($n_{is_i}=n_{js_j}$ only if $s_i\neq s_j$). The total number of 
excitation should be taken the flavor into account, thus 
\begin{equation}
d = \sum_{i,s} \delta(n_{is} -1).
\end{equation}

\begin{figure}
	\centering
	\includegraphics[width=0.7\textwidth]{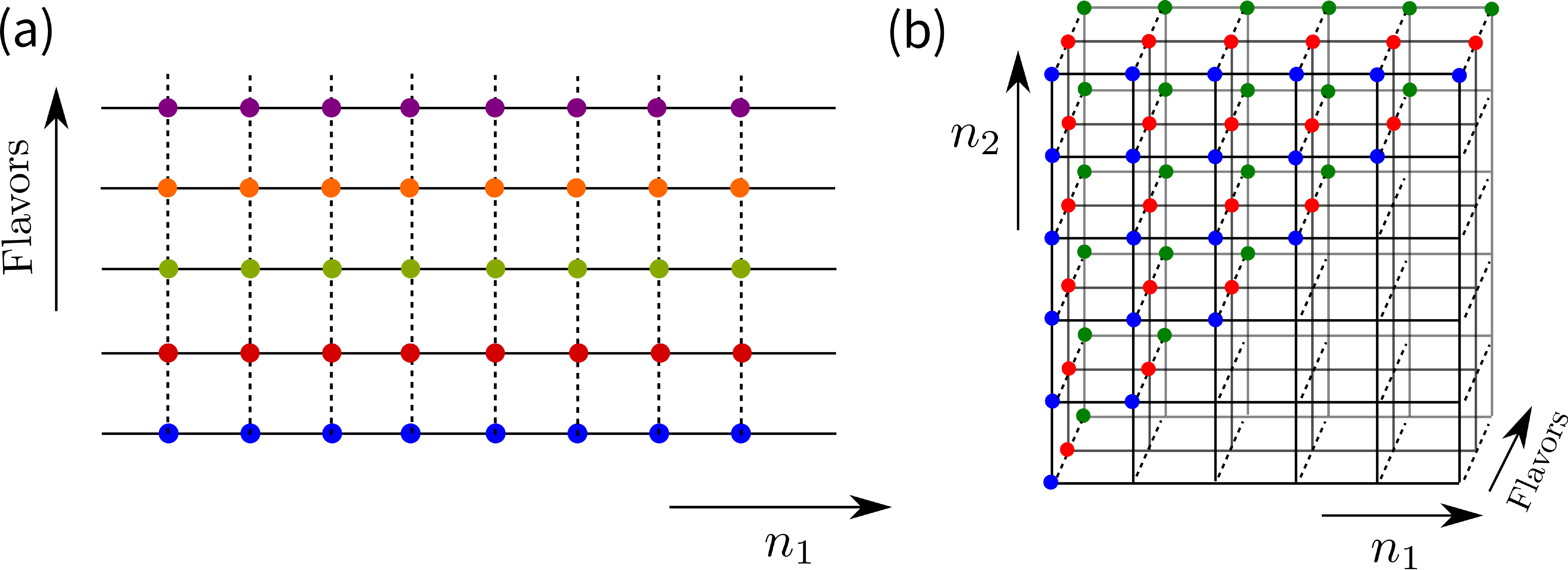}
	\caption{{\bf Flavors in MBL for spinful fermions and bosons}. Example for (a) 1-dimensional, (b) 2-dimensional virtual lattice. In each site, the spin degree of freedom in the Fermi-Hubbard model or Bose-Hubbard model can be regarded as some kind of flavor. In this way, each site in the $d$-dimensional virtual lattice can contain structure, which does not influence our basic picture for MBL.}
	\label{fig-figsup11}
\end{figure}

\subsection{B: Bose-Hubbard model with the conserved number of particle}

The most widely explored 1D spinless Bose-Hubbard model can be written as \cite{andraschko2014purification}  
\begin{equation}
	H = -t\sum_i^L \hat{b}_i^\dagger \hat{b}_{i+1} +H.c. + \frac{U}{2} \hat{n}_i (\hat{n}_i -1) + v_i \hat{n}_i,
\end{equation}
where $\hat{n}_i = \hat{b}_i^\dagger \hat{b}_i$. Let us denote the basis in the same way as Eq. \ref{eq-phin} with $1\leq n_1\leq n_2\leq n_3\leq \cdots \leq n_d\leq L$, where $n_i$ represents the creation of a boson at the $n_i$-th site and $d$ is the total number of particles.  
We see that except the boundary condition added $n_i=n_j$, all the other issues are exactly the same as that for the spin model.

However, for a spinful Bose-Hubbard 
\begin{equation}
	H = -t\sum_i^L \hat{b}_{i,s}^\dagger \hat{b}_{i+1,s} + H.c. + \frac{U_{i,s}}{2} \hat{n}_{i,s} (\hat{n}_{i,s} -1) + v_{i,s} \hat{n}_{i,s},
\end{equation}
where $\hat{n}_{i,s} = \hat{b}_{i,s}^\dagger \hat{b}_{i,s}$. Similar to the spinful fermion case, the basis can be defined in Eq. \ref{eq-SpinFermi} and the spin degree of freedom can
be denoted as some kind of flavor in each site, which shows in Fig. \ref{fig-figsup11}.  

We can conclude that our basic picture of 
MBL as some kind of AL in the virtual lattice with the correlated disorder should be true even for spinful fermions and bosons.

\subsection{C: Effect of random interaction and the equivalent long-range correlated random potential}

Here we want to show that in the presence of random on-site many-body interaction, instead of random on-site disorder potential, will yield the same physics. To this end,
we consider the XXZ model with periodic boundary condition in the following way,
\begin{equation}                                                                                                                                                                    H = \sum_i J(S_i^x S_{i+1}^x + S_i^y S_{i+1}^y) + J_z^{i} S_i^z S_{i+1}^z + h\sum_i S_i^z,
	\label{eq-XXZ}
\end{equation}
where $J_z^i$ is some kind of random interaction between $i$-th site and $(i+1)$-th site. It is well-known that this term, after Jordan-Wigner transformation, will be reduced to the fermion model with many-body interaction.  Following the main text, the wave function with $d$ spin excitation can be defined in Eq. 4. With the same method, The equivalent single particle tight-binding Hamiltonian for Eq. \ref{eq-XXZ} is given by
\begin{equation}
H =  \sum_{\langle\textbf{n},\textbf{n}^\prime\rangle} \frac{J}{2} (\hat{c}_{\textbf{n}}^\dagger \hat{c}_{\textbf{n}^\prime} + h.c.) 
 +\sum_{\textbf{n}} [ -\frac{h}{2}(L-2d) + \eta + \xi_{\textbf{n}} ] \hat{c}_{\textbf{n}}^\dagger \hat{c}_{\textbf{n}},
	\label{eq-HdRI}
\end{equation}
where $\eta = \sum_i^L \frac{J_z^i}{4}$ and the random potential
\begin{equation}
 \xi_{\textbf{n}} = -\frac{1}{2}(J_z^{n_1-1}+ J_z^{n_d}) - \frac{1}{2} \sum_i^{d-1} [(1-\delta_{m_i,1})(J_z^{n_i} + J_z^{n_{i+1}-1})], \qquad m_i = n_{i+1}-n_i.
\end{equation} 
If we let $J_z^i \in[-W,W]$, for two sites ($\mathbf{n}\neq\mathbf{n}^\prime$), the correlation of the random potential is given by
\begin{equation}
	 \langle \xi_{\mathbf{n}} \xi_{\mathbf{n}^\prime} \rangle =  \langle   \frac{1}{4} \sum_{i,j}(J_z^{n_i-1}J_z^{n_j^\prime-1} + J_z^{n_i-1}J_z^{n_j^\prime} +J_z^{n_i}J_z^{n_j^\prime-1} + J_z^{n_i}J_z^{n_j^\prime}) \rangle =  \langle \frac{1}{4}  \sum_{\mathcal{D}}^k (J_z^{n_i})^2 \rangle = \frac{kW^2}{12},
	\label{eq-correlation}
\end{equation} 
where $\mathcal{D}\equiv [(n_i=n_j^\prime)$ or $(n_i=n_j^\prime-1)$ or $(n_i-1=n_j^\prime)]$ and $k \le d$. This means that the disordered many-body interaction also introduces long-range correlated disorder in the virtual lattice. However, these two cases have an important difference. 
In the model with random on-site potential, AL and MBL may co-exist; while in the case with random many-body interaction, the single-particle AL (for $d = 1$) is 
strictly absent. This observation is suggestive that AL and MBL may be two totally different concepts, as discussed in the introduction of the main text.

\subsection{D: XYZ model with $\mathbb{Z}_2$ symmetry} 

In the above discussion, $S_\text{z}$ in the XXZ model and the total number of particles in the fermion and boson models are conserved quantities, which divide the whole Hilbert space into
different subspaces. In this case, the dimension $d$ is well defined. In the following XYZ model, 
\begin{equation}
    H = \sum_i J_x S_i^x S_{i+1}^x + J_y S_i^y S_{i+1}^y + J_z S_i^z S_{i+1}^z + h_i S_i^z,
	\label{eq-XYZ}
\end{equation}
the $S_\text{z}$ is not commuted with $H$, thus it is not a good quantum number. In this case, the good quantum number is characterized by 
\begin{equation}
	P_\text{z} = \prod_i S_i^z,
\end{equation}
thus $P_\text{z}^2 = 1$ and $P_\text{z} = \pm 1$. This $\mathbb{Z}_2$ symmetry is reflected from the fact that $S_i^z \rightarrow S_i^z$ 
and  $S_i^{x,y} \rightarrow -S_i^{x,y}$ for all $i$. In this case, the subspaces defined in the XXZ model can still be defined, however,
they will be coupled by the XYZ model. In all these subspaces, the cases with $d \simeq L/2$ have the largest weight (noticed that
from the Stirling formula, $C_L^d$ will approach a Gaussian distribution centered at $d = L/2$, with variance $\sigma^2=L$). 
For this reason, the MBL happened in this case can still be regarded as an infinity dimensional AL. 

Let us briefly mention that after a Jordan-Wigner transformation, this model is reduced to the superconducting model, where $J_z$ 
is reduced to the interaction term. In this case, the above $\mathbb{Z}_2$ symmetry is reduced to the parity symmetry with even 
number of particles and odd number of particles. The similar physics happens to the boson models. For this reason, even in the case 
with other symmetries, the MBL phase can still be interpreted as some kind of AL in infinite-dimensional. 

\section{S6: Different from our definition of MBL and the previous model with Fock state in the Bethe lattice}

\begin{figure}
	\centering
	\includegraphics[width=0.6\textwidth]{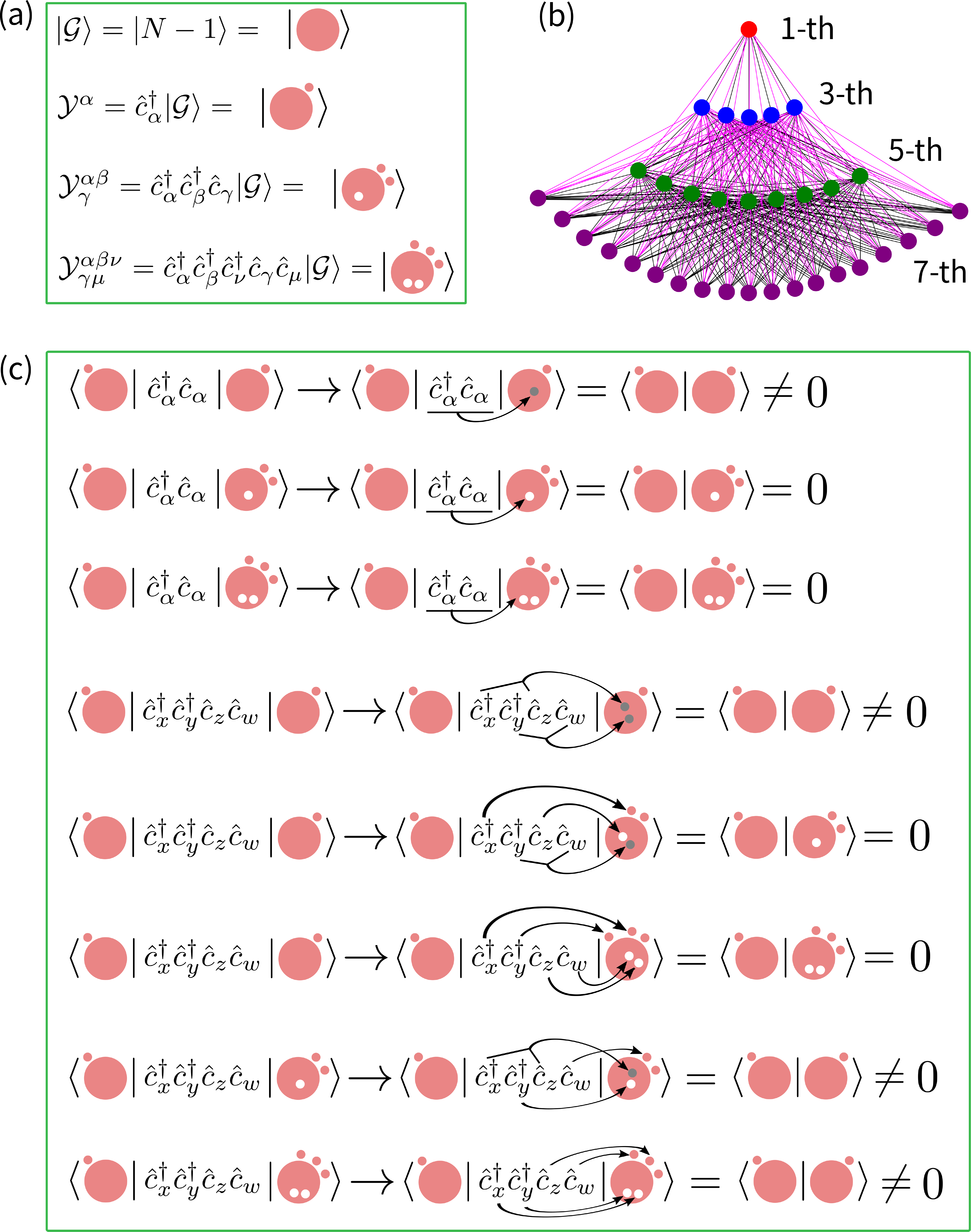}
	\caption{{\bf Fock space configuration and lattice constructed by these generations}. (a) Fock states: the ground state of $N-1$ particles is defined by $|\mathcal{G}\rangle$; the 1-th, 3-th and 5-th generation are given by $\mathcal{Y}^\alpha$, $\mathcal{Y}^{\alpha\beta}_{\gamma}$ and $\mathcal{Y}^{\alpha\beta\nu}_{\gamma\mu}$. (b) Lattice formed by these generations due to coupling induced by the many-body interaction. In some limiting case, this complicated lattice structure can be reduced to the cycle-free Bethe lattice (see main text). (c) Effective hopping between different generations and the same generations give on-site potential. Any state from generation $2n-1$ is connected to states from $2n-5$, $2n-3$, $2n-1$, $2n+1$, or $2n+3$. }
	\label{fig-figsup12}
\end{figure}

In this section, we aim to discuss the relation between our model and the previous definition of MBL in the lattice constructed by the Fock states,
which in some limiting cases can be reduced to the cycle-free Bethe lattice. Our new definition of MBL can be regarded as some kind of conceptional 
extension of the previous model by placing the Fock space in a $d$ dimensional virtual lattice. 

In the previous section, we have shown that our picture for MBL is applicable to spin, fermion and boson models. Here we consider the following 
many-body model, following the discussion in Ref. \cite{altshuler1997quasiparticle}
\begin{equation}
H = \sum_\alpha \varepsilon_\alpha \hat{c}_\alpha^\dagger \hat{c}_\alpha + \sum_{xyzw} V_{xyzw} \hat{c}_x^\dagger \hat{c}_y^\dagger \hat{c}_z \hat{c}_w.
\end{equation}
Let $|\mathcal{G}\rangle = |N-1\rangle$ to be the ground state of $N-1$ particles, then the Fock states can be constructed using
\begin{equation}
\mathcal{Y}^{i_1,i_2,\cdots,i_n}_{j_1,j_2,\cdots,j_{n-1}} = \hat{c}_{i_1}^\dagger \hat{c}_{i_2}^\dagger \cdots \hat{c}_{i_n}^\dagger \hat{c}_{j_1} \hat{c}_{j_2} \cdots \hat{c}_{j_{n-1}}|\mathcal{G}\rangle,
\end{equation}
which represents $n$ particles and $n-1$ hole, form the $2n-1$ generation, which are shown in Fig. \ref{fig-figsup12} (a). In these basis, the localization occurs in the Fock space of many-body states, 
rather than in the real space. 
Both single-particle and two-body interaction can be introduced into the on-site potential. By the two-body interaction, any state from generation $2n-1$ is connected to states from $2n-5$, $2n-3$, $2n-1$, $2n+1$, or $2n+3$, which is shown in Fig. \ref{fig-figsup12} (c). The lattice in the Fock space of many-body states shows in Fig. \ref{fig-figsup12} (b). However, in Ref. \cite{altshuler1997quasiparticle}, the coupling between the next nearest neighbors are ignored, which means that $V_{xyzw}=0$ for $x\neq y\neq z\neq w$ and any state from generation $2n-1$ is connected to states from  $2n-3$, $2n-1$, or $2n+1$. In this limit this complicated lattice (see Fig. \ref{fig-figsup12} (b)) can be reduced to the cycle-free Bethe lattice (BL). 

We find that the above picture is different from our picture for MBL, though a similar basis is used. In the BL lattice model, the Fock space is 
represented by a single node characterized by different generations, while the structure of these nodes is not specified. 
However, in our model, it is characterized by the nodes in the $d$ dimensional lattice subjected to some proper boundary 
condition. We show in Fig. \ref{fig-figsup7} that when $L \rightarrow \infty$, the dimension $D_0 \rightarrow d$, thus the boundary effect may be 
neglected. Moreover, in our lattice model, no approximation is required, which naturally yields nearest neighboring interaction (see Fig. 1 (b) and 
Fig. 1 (c) in the main text), thus can be much more easily formulated during the numerical simulation. In our model, the dimension $d$ has specific 
physical meaning, thus from the conceptional point of view, our lattice model is much more transparent than the BL lattice based on 
different "generations". In our model, the construction of $|\mathcal{G}\rangle$ is not necessary. Since each node has definite physical meaning, 
this $d$-dimensional virtual lattice may also be useful to understand the evolution of entanglement entropy, which may contain the contribution from 
both the configurational entanglement entropy and number entanglement entropy \cite{lukin2019probing}. This new definition of MBL may also be useful
to understand the absence of mobility edge in the experiment \cite{kohlert2019observation, abanin2019many}.
\end{widetext}

\end{document}